\PassOptionsToPackage{pdfpagelabels=false}{hyperref}
\documentclass[fleqn,usenatbib,useAMS]{mnras}
\usepackage{amsmath}
\usepackage{amssymb}

\usepackage{graphicx}
\usepackage[figure,figure*,table,table*]{hypcap}
\usepackage{multirow}
\usepackage{algorithm}
\usepackage{algpseudocode}

\title[\nautilus{}: boosting importance nested sampling]{\nautilus{}: boosting Bayesian importance nested sampling with deep learning}
\author[J. U. Lange]{Johannes U. Lange$^{1, 2, 3, 4}$\thanks{email: julange.astro@pm.me}\\
$^1$Kavli Institute for Particle Astrophysics and Cosmology and Department of Physics, Stanford University, CA 94305, USA\\
$^2$Department of Astronomy and Astrophysics, University of California, Santa Cruz, CA 95064, USA\\
$^3$Department of Physics, University of Michigan, Ann Arbor, MI 48109, USA\\
$^4$Leinweber Center for Theoretical Physics, University of Michigan, Ann Arbor, MI 48109, USA}
\pubyear{2022}

\newcommand{\nautilus}{{\sc nautilus}}
\newcommand{\emcee}{{\sc emcee}}
\newcommand{\pocomc}{{\sc pocoMC}}
\newcommand{\ultranest}{{\sc UltraNest}}
\newcommand{\dynesty}{{\sc dynesty}}
\newcommand{\multinest}{{\sc MultiNest}}

\begin{document}

\date{Accepted xxx. Received xxx}

\label{firstpage}
\pagerange{\pageref{firstpage}--\pageref{lastpage}}

\maketitle

\begin{abstract}
    We introduce a novel approach to boost the efficiency of the importance nested sampling (INS) technique for Bayesian posterior and evidence estimation using deep learning. Unlike rejection-based sampling methods such as vanilla nested sampling (NS) or Markov chain Monte Carlo (MCMC) algorithms, importance sampling techniques can use all likelihood evaluations for posterior and evidence estimation. However, for efficient importance sampling, one needs proposal distributions that closely mimic the posterior distributions. We show how to combine INS with deep learning via neural network regression to accomplish this task. We also introduce \nautilus{}, a reference open-source Python implementation of this technique for Bayesian posterior and evidence estimation. We compare \nautilus{} against popular NS and MCMC packages, including \emcee{}, \dynesty{}, \ultranest{} and \pocomc{}, on a variety of challenging synthetic problems and real-world applications in exoplanet detection, galaxy SED fitting and cosmology. In all applications, the sampling efficiency of \nautilus{} is substantially higher than that of all other samplers, often by more than an order of magnitude. Simultaneously, \nautilus{} delivers highly accurate results and needs fewer likelihood evaluations than all other samplers tested. We also show that \nautilus{} has good scaling with the dimensionality of the likelihood and is easily parallelizable to many CPUs.
\end{abstract}


\section{Introduction}

Modern astronomy and cosmology rely heavily on Bayesian inference. This allows us to estimate or update our belief in the probability distribution of model parameters in the presence of experimental data. Using Bayes' theorem, we can calculate the probability density of the model parameters $\mathbf{\Theta}$ given observational data $D$ via
\begin{equation}
    P(\mathbf{\Theta} | D) = \frac{\mathcal{L} (D | \mathbf{\Theta}) \pi(\mathbf{\Theta})}{\mathcal{Z}(D)} \, .
\end{equation}
In the above equation, $P(\mathbf{\Theta} | D)$ is the so-called posterior distribution of model parameters. Similarly, $\mathcal{L} (D | \mathbf{\Theta})$ is the likelihood, the probability density of observing $D$ given model parameters $\mathbf{\Theta}$. $\pi(\mathbf{\Theta})$ is the prior probability and represents our prior knowledge or expectation of model parameters. Finally, $\mathcal{Z}(D)$ is the Bayesian model evidence and represents the probability of obtaining $D$, averaged over all model parameters. Given the normalisation of the posterior probability, $\int P(\mathbf{\Theta} | D) d\mathbf{\Theta} = 1$, the Bayesian evidence is computed via
\begin{equation}
    \mathcal{Z}(D) = \int \mathcal{L}(D | \mathbf{\Theta}) \pi(\mathbf{\Theta}) d\mathbf{\Theta} \, .
    \label{eq:evidence}
\end{equation}
In the context of Bayesian inference, one may estimate parameter expectation values such as the means and uncertainties of individual model parameters $\theta_i$. Similarly, Bayesian model comparison involves estimating the normalizing constant, the Bayesian evidence $\mathcal{Z}(D)$. In both cases, this involves multi-dimensional integrals over the posterior $P(\mathbf{\Theta} | D)$ and is often practically impossible with standard integration methods. For traditional numerical methods such as Gaussian quadrature, the number of likelihood evaluations scales as $\mathcal{O}(\exp(N_{\rm dim}))$, where $N_{\rm dim}$ is the number of model parameters or dimensions, and can quickly become unfeasible. Instead, Bayesian analysis methods often rely on Monte-Carlo methods such as Markov chain Monte Carlo (MCMC), with \emcee{} \citep{ForemanMackey2013_PASP_125_306} being one of the most widely-used implementations in astronomy. In MCMC, one generates a sequence, a Markov chain, of samples of $\mathbf{\Theta}$ that, on average, are distributed proportionally to $P(\mathbf{\Theta} | D)$. One can then estimate posterior quantities, such as one-dimensional parameter uncertainties, using samples from the Markov chain. While MCMC breaks the steep dimensionality scaling of standard integration methods, it is still computationally expensive, often requiring millions of evaluations of $\mathcal{L}(D | \mathbf{\Theta})$. Furthermore, MCMC may require hand-tuning algorithm hyper-parameters, and results must be carefully checked for convergence. Finally, MCMC by itself cannot be used to estimate the Bayesian evidence, and it struggles with multi-modal distributions with widely separated peaks.

Especially in astronomy and cosmology, nested sampling \citep[NS; ][]{Skilling2004_AIPC_735_395} has emerged as one of the most widely-used alternatives to MCMC. In NS, we start by randomly drawing a large number $N_{\rm live}$ of ``points'' from the entire prior set and evaluate their likelihood. These points form the so-called live set. Afterwards, we iteratively update the live set by drawing new samples from the prior. Whenever we draw a new point with a likelihood higher than the lowest value in the live set, $\mathcal{L}_{\rm min}$, we replace the lowest-likelihood point with the newly drawn one. Points that left the live set constitute the inactive set. By construction, the live set will contain points of ever-increasing likelihood. The NS algorithm stops once the likelihood does not improve substantially anymore. One can show that the volume that the live set represents shrinks, on average, exponentially each time a point in the live set is replaced. Using this property, one can combine live and inactivate sets to produce samples from the posterior and estimate the Bayesian evidence \citep{Skilling2004_AIPC_735_395}. Popular implementations of NS include \multinest{} \citep{Feroz2009_MNRAS_398_1601}, {\sc PolyChord} \citep{Handley2015_MNRAS_453_4384} and and \dynesty{} \citep{Speagle2020_MNRAS_493_3132}.

In recent years, there has been increased interest in using deep learning based on neural networks to improve Bayesian Monte-Carlo methods \citep[see, e.g.][]{Jia2019_arXiv_1912_6073, Moss2020_MNRAS_496_328, Karamanis2022_MNRAS_516_1644, To2023_JCAP_01_016}. Following this idea, in this work, we present a new Bayesian inference algorithm based on combining importance nested sampling \citep[INS;][]{Feroz2019_OJAp_2_10}, an extension of traditional NS, with neural networks for efficient proposals. We show that this new algorithm has several significant benefits over popular Bayesian inference codes.

The outline of the paper is as follows. In section \ref{sec:ins}, we describe the INS algorithm and also introduce improvements over the original algorithm described in \cite{Feroz2019_OJAp_2_10}. In section \ref{sec:bounds}, we explain how we can use neural networks to boost the sampling efficiency of the algorithm further. Afterwards, in section \ref{sec:application}, we test our algorithm as implemented in the \nautilus{} Python package against various established Bayesian codes. Finally, we discuss our findings in section \ref{sec:discussion} and conclude in \ref{sec:conclusion}. Throughout this work, we use $\log$ to denote the natural logarithm.

\section{Importance nested sampling}
\label{sec:ins}

Within Bayesian posterior and evidence estimation, we typically want to achieve several goals. First, we want to estimate the region of parameter space $\mathbf{\Theta}$ that exhibits the highest likelihood $\mathcal{L} (D | \mathbf{\Theta})$. Second, we want to draw random samples proportional to the posterior $P(\mathbf{\Theta} | D) \propto \mathcal{L} (D | \mathbf{\Theta}) \pi(\mathbf{\Theta})$. And third, we may want to calculate the Bayesian evidence $\mathcal{Z} (D)$ for Bayesian model comparison.

Naive implementations of the NS algorithm would suffer from an extremely low sampling efficiency: as the volume of the live set shrinks exponentially, so does the fractional volume compared to the prior. Always drawing new points from the entire prior would quickly result in virtually no new live points. Practical implementations of the NS algorithm such as \multinest{} \citep{Feroz2009_MNRAS_398_1601} thus typically use information about the live set to make proposals from a much smaller bounding sub-volume around the live points, keeping sampling efficiency high as the live set volume decreases. However, the expectation that the volume of the live set decreases exponentially is only valid if new points are proposed uniformly from the entire live set volume. In other words, the results of the NS algorithm would be biased if the proposal volume does not fully encompass the so-called iso-likelihood surface, the set of points for which $\mathcal{L} = \mathcal{L}_{\rm min}$. Given the difficulty of estimating such volumes in high-dimensional spaces, NS implementations typically weigh accuracy against sampling efficiency. If the proposal volume for new points is larger (smaller), one misses less (more) of the iso-likelihood surface while having a smaller (larger) fraction of points accepted. We refer the reader to \citet{Ashton2022_NRvMP_2_39} for a detailed review of the NS algorithm and its implementations.

More recently, \cite{Feroz2019_OJAp_2_10} introduced a variation of the NS algorithm called INS. In essence, instead of relying on the expectation that the live set volume shrinks exponentially, we use the fact that with certain sampling strategies, so-called region samplers, we can very accurately estimate the distribution from which new points are proposed. Using this distribution as a pseudo-importance function, one then assigns a weight to all the points drawn, whether they ever make it into the live set or not. Using this insight, compared to the original NS algorithm, the INS algorithm can use all points for which the likelihood was evaluated to estimate the posterior and evidence, not just the small fraction that made it into the live set. Another advantage of INS is that it is not necessary for the proposal volume to fully encompass the iso-likelihood surface for the results to be accurate. Thus, the INS algorithm has the potential to produce more accurate results than practical implementations of the NS algorithm. The following gives an overview of the INS algorithm implemented in this work. Although similar, we note a few essential differences between the INS algorithm described here and the one outlined in \cite{Feroz2019_OJAp_2_10}. This includes the ability to draw additional samples after the initial algorithm terminated, making our version a ``dynamic'' INS algorithm in analogy to dynamic NS algorithms \citep{Higson2019_SC_29_891}. As a result, this new algorithm has two phases: an exploration phase and an optional sampling phase. In this section, we start by reviewing the aspect of importance sampling in the proposed INS algorithm before describing the two phases.

\subsection{Importance sampling}

\begin{figure*}
    \centering
    \includegraphics{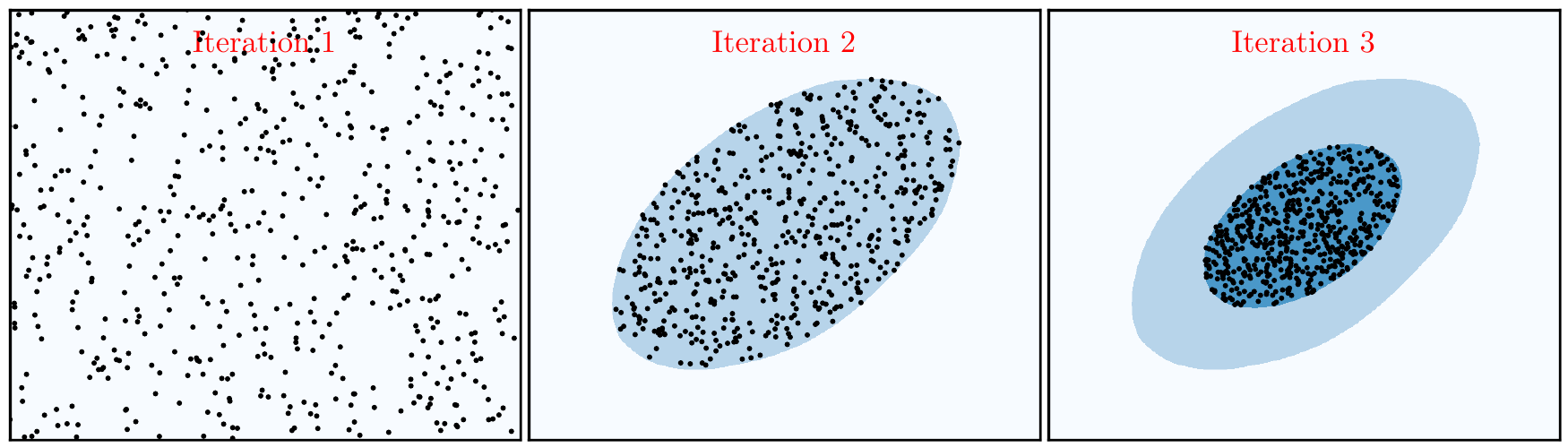}
    \caption{Schematic view of the exploration phase of the INS algorithm. We start by sampling the entire prior uniformly with random points. Then, the points with the highest likelihood are identified, and the volume they represent is sampled further. This step is repeated until a convergence criterion is met.}
    \label{fig:exploration}
\end{figure*}

\begin{figure*}
    \centering
    \includegraphics{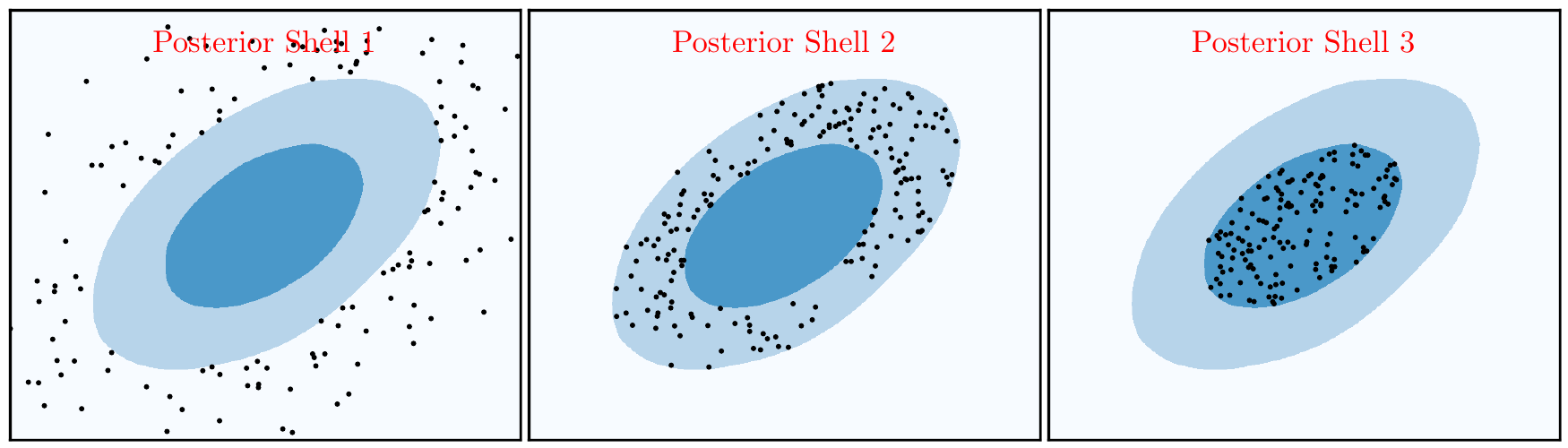}
    \caption{Schematic view of how the INS algorithm estimates the posterior. For each shell, the volume is sampled uniformly so we can assign a corresponding sampling density $g$. We then assign an importance weight to each point proportional to the ratio of its likelihood $\mathcal{L}$ to the sampling density $g$. The total posterior is then estimated by summing the contributions from all shells. Although we receive a weighted posterior sample, for visualization purposes, we downsampled points in the above figure according to the importance weight to receive an equal-weighted posterior sample.}
    \label{fig:posterior}
\end{figure*}

During the exploration phase, we sample points from exponentially shrinking bounding volumes $B_i$. A schematic view of this approach is shown in Fig.~\ref{fig:exploration}. Let us define the $i$-th shell as the difference of $B_i$ and all subsequent bounds $B_k$,
\begin{equation}
    S_i = B_i \setminus \left( \cup_{k=i+1}^{N_{\rm shells}} B_k \right) \, .
\end{equation}
When defined this way, no shells overlap, and the union of all shells represents the entire prior space. Thus, each point in the prior can be uniquely associated with a shell. If each shell is uniformly sampled, we can define a pseudo-importance sampling density $g$ via
\begin{equation}
    g(\mathbf{\Theta}) = \frac{N_i}{V_i} \,
    \label{eq:pseudo-density}
\end{equation}
where $V_i$ corresponds to the volume of the $i$-th shell and $N_i$ is the corresponding number of points sampled. While the points evaluated by the INS algorithm sample $g$, we can make them sample the posterior through an importance weighting of each point,
\begin{equation}
    w = \frac{\mathcal{L} (D | \mathbf{\Theta}) \pi(\mathbf{\Theta})}{g(\mathbf{\Theta})} \, .
    \label{eq:weight_per_point}
\end{equation}
Finally, the estimate for the evidence $\mathcal{Z}$ has a similarly simple estimator,
\begin{equation}
    \mathcal{Z} (D) = \sum_k w_k \, .
    \label{eq:evidence_estimate}
\end{equation}
Fig.~\ref{fig:posterior} gives a schematic view of how the results from different shells are combined to estimate the posterior. During the sampling phase, we can sample each shell further, i.e., add more points, to enhance our estimates of the posterior and the evidence. Even more, we can do so dynamically, adding points to those shells that add the largest uncertainties to our posterior and evidence estimates.

We note that the original INS algorithm presented in \citet{Feroz2019_OJAp_2_10} and implemented in \multinest{} assumed strictly shrinking bounding volumes, i.e., $B_i \subset B_k$ for all $i > k$. This allowed the algorithm not to store information about all bounding volumes, only the most recent one. However, the above assumption is unlikely to be correct in non-trivial INS applications. The boundaries that \multinest{} draws are based on multi-ellipsoidal decomposition and may partially reach outside of previous bounds. We speculate that this may be partially responsible for some of the apparent biases in the INS results of \multinest{} reported in the literature \citep{Lemos2023_MNRAS_521_1184} and in seen in other tests conducted by the author.

\subsection{Exploration phase}

In line with NS algorithms, we first assume that the prior space $\mathbf{\Theta} = (\theta_1, \theta_2, ..., \theta_n)$ is an $n$-dimensional hypercube, i.e., $0 \leq \theta_i < 1$ for all $1 \leq i \leq n$, and that the prior on $\mathbf{\Theta}$ is flat, i.e., $\pi(\mathbf{\Theta}) \equiv 1$. While this is not true for general priors, in many applications, i.e., those where the prior is separable, one can achieve this via a prior transform where $\theta_i$ is transformed into the value of the cumulative distribution function (CDF) at $\theta_i$,
\begin{equation}
    \theta_i = \int\limits_{-\infty}^{\theta_i} \pi(\tilde{\theta}_i) d \tilde{\theta}_i \, .
\end{equation}
By definition, the transformed parameter values are in the unit range and have a flat prior. We note that not all priors may be separable. We refer the reader to \cite{Alsing2021_MNRAS_505_95} for a detailed discussion of the prior transform for the NS algorithm and how to implement arbitrary prior distributions using parametric bijectors. Additionally, in appendix~\ref{sec:arbitrary_prior}, we describe a simple alternative approach of implementing non-separable priors by absorbing the prior into the likelihood \citep{Feroz2009_MNRAS_398_1601}.

\begin{algorithm}
    \begin{algorithmic}[1]
    \State $B \gets$ Hypercube
    \State $\mathcal{L}_{\rm min} \gets 0$
    \State live set $\gets$ $N_{\rm live}$ points from $B$
    \While{$f_{\rm live} > f_{\rm live, max}$}
        \State $N \gets 0$
        \While{$N < N_{\rm update}$}
            \State draw point $\mathbf{\Theta}$ from $B$
            \If{$\mathcal{L}(\mathbf{\Theta}) > \mathcal{L}{\rm min}$}
                \State $N \gets N + 1$
            \EndIf
        \EndWhile
        \State live set $\gets$ $N_{\rm live}$ points with highest $\mathcal{L}(\mathbf{\Theta})$
        \State $B \gets$ new bound around live set
        \State $\mathcal{L}_{\rm min} \gets$ lowest $\mathcal{L}(\mathbf{\Theta})$ of live set
    \EndWhile
    \end{algorithmic}
    \caption{Exploration Phase}
    \label{alg:exploration}
\end{algorithm}

Algorithm \ref{alg:exploration} gives a schematic view of the exploration phase of the INS algorithm. The two important parameters during that phase are $N_{\rm live}$, the number of live points, and $N_{\rm update}$, the number of ``updates'' before a new bound is drawn. Here, we define an update as drawing a point with a likelihood higher than the lowest likelihood of the live set used to draw the current bound. At the beginning of the INS algorithm, the bound we draw from is the entire prior volume, the $n$-dimensional hypercube. Without prior knowledge about the likelihood, the corresponding minimum likelihood we seek to achieve is $\mathcal{L}_{\rm min} = 0$. Thus, at the beginning of the INS algorithm, we draw $N_{\rm live} + N_{\rm update}$ points sampled uniformly from the unit hypercube. The first $N_{\rm live}$ points initiate the live set, and drawing $N_{\rm update}$ points leads to $N_{\rm update}$ updates since any likelihood should be positive. Thus, we have successfully filled the first bound, the unit hypercube. We now define the new live set as the $N_{\rm live}$ points with the highest likelihood out of the $N_{\rm live} + N_{\rm update}$ points we already drew. We then define a new sub-volume of the entire prior space that encompasses all or most points of the live set and, ideally, only a few points that did not make it into the current live set. The lowest likelihood value we seek to achieve is the lowest likelihood of the $N_{\rm live}$ points, $\mathcal{L}_{\rm min}$. With the new bound defined, we draw points from this new bound until $N_{\rm update}$ points have been drawn whose likelihood is larger than $\mathcal{L}_{\rm min}$ at which point we build another bound centred on the $N_{\rm live}$ points with the highest likelihood. We then repeat this procedure of drawing and filling new bounds until the convergence criterion is met. We follow other NS algorithms and base the stopping criterion on the evidence $\mathcal{Z}_{\rm live}$ in the live set, which can be estimated by summing the weights of all live points, in analogy to eq.~\eqref{eq:evidence_estimate}. By default, we will stop the algorithm once $f_{\rm live}$, defined via
\begin{equation}
    f_{\rm live} = \mathcal{Z}_{\rm live} / \mathcal{Z}
\end{equation}
drops below $0.01$, i.e., less than $1\%$ of the evidence remaining in the live set.

Note that our estimates of the posterior and the evidence implicitly assume that each shell is uniformly sampled by the points drawn in the INS algorithm. We note that at each iteration $i$, newly drawn points are sampled uniformly from bound $B_i$. However, it is not guaranteed that previously drawn points falling into $B_i$ sample it uniformly. In appendix \ref{sec:uniform_sampling}, we describe this problem and how to solve it.

\subsection{Sampling phase}
\label{subsec:sampling_phase}

\begin{algorithm}
    \begin{algorithmic}[1]
    \While{$N_{\rm eff} \leq N_{\rm eff, min}$}
        \State $i \gets \arg\max_i \mathcal{Z}_i / (N_{{\rm eff}, i} N_i)^{1/2}$
        \State draw point $\mathbf{\Theta}$ from $S_i$
        \State $N_i \gets N_i + 1$
        \State update $N_{\rm eff}$, $N_{{\rm eff}, i}$ and $\mathcal{Z}_i$
    \EndWhile
    \end{algorithmic}
    \caption{Sampling Phase}
    \label{alg:sampling}
\end{algorithm}

Once the exploration phase is completed, we can add more points to individual shells, thereby increasing the precision of our posterior and evidence estimates. We aim to increase the effective sample size, $N_{\rm eff}$. This quantity is defined via
\begin{equation}
    N_{\rm eff} = \frac{\left[ \sum w_k \right]^2}{\sum w_k^2} \, ,
    \label{eq:ess}
\end{equation}
where the sum goes over all points and the weight assigned to each point is defined in eq.~\eqref{eq:weight_per_point}. We can estimate the evidence associated with each shell,
\begin{equation}
    \mathcal{Z}_i = V_i \langle \mathcal{L} \rangle_i = V_i \frac{\sum_{k=1}^{N_i} \mathcal{L}_k}{N_i} \, .
    \label{eq:evidence_per_shell}
\end{equation}
as well as a corresponding effective sample size $N_{{\rm eff}, i}$. Let us denote the sampling efficiency as
\begin{equation}
    \eta_i = N_{{\rm eff}, i} / N_i \, ,
\end{equation}
which can be defined for individual shells, as above, and analogously for the combination of all shells. An optimal sampling strategy, one that maximises $\eta$, is to sample each shell proportionally to $\mathcal{Z}_i \eta_i^{-1/2}$. This implies that we add new points to the shell for which $\mathcal{Z}_i \eta_i^{-1/2} N_i^{-1}$ is maximal, as outlined in Algorithm \ref{alg:sampling}. Under this strategy, the total sampling efficiency becomes $\eta = \left[ \sum \eta_i^{-1/2} (\mathcal{Z}_i / \mathcal{Z}) \right]^{-2}$. By default, we choose a target of $N_{\rm eff} > 10,000$. One can show that the relative uncertainty in the evidence, $\Delta \log \mathcal{Z}$, is roughly $N_{\rm eff}^{-1/2}$, i.e., by default, we aim for $\Delta \log \mathcal{Z} \approx 0.01$. As we will explore later, sampling from the posterior at the sampling stage can be very efficient, i.e., the effective sample size increases roughly at the same speed as the number of likelihood evaluations, $\eta \sim 0.1 - 1$.

Finally, note that we call $g$, as defined in eq.~\eqref{eq:pseudo-density}, a \textit{pseudo}-importance sampling density. This is because $g$ is defined a posteriori when constructing each new bound during the exploration phase, subtly violating the implicit assumption that points are sampled randomly from each shell. This fact can lead to small biases in the posterior and evidence estimates, which we will explore later. However, this issue can be overcome by discarding points drawn in the exploration phase and drawing new ones during the sampling phase. This is somewhat akin to discarding the so-called burn-in phase in MCMC, though we will see that the pseudo importance bias is often negligible in practice.

\section{Neural Network Bounds}
\label{sec:bounds}

\begin{figure*}
    \centering
    \includegraphics[width=\textwidth]{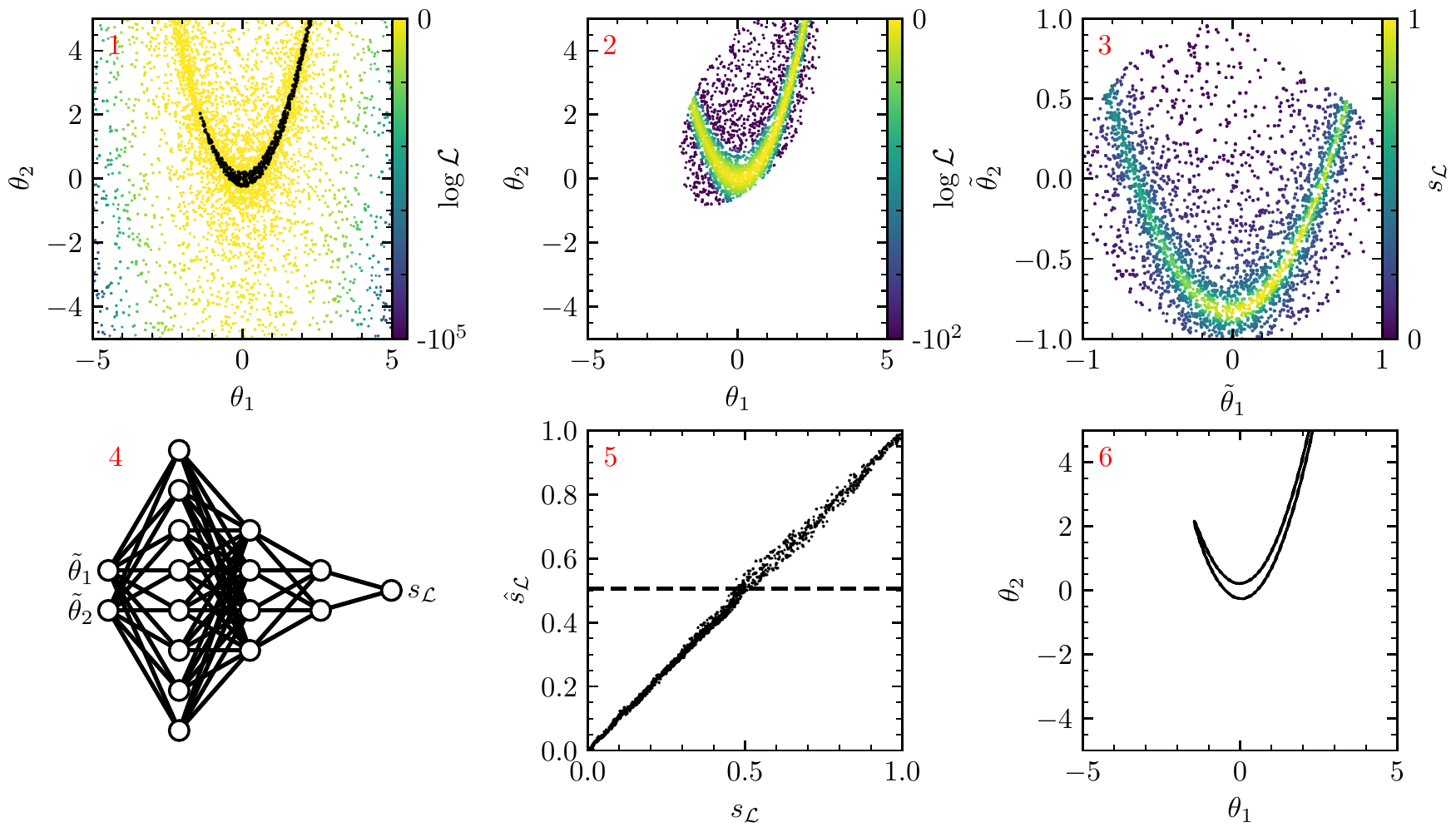}
    \caption{Diagram depicting how new proposal volumes during the exploration phase are constructed. For this example, we chose the two-dimensional Rosenbrock likelihood. The steps are as follows. (1) The set of $N_{\rm live}$ points with the highest likelihood, the so-called live set, is identified. (2) One or multiple non-overlapping bounding ellipsoids are drawn around the live set. (3) The coordinates $\mathbf{\Theta}$ of points in the ellipsoid are transformed into the ellipsoid coordinates $\mathbf{\tilde{\Theta}}$ using a Cholesky decomposition. Similarly, likelihood values are converted into likelihood scores $0 \leq s_{\mathcal{L}} \leq 1$. (4) The transformed coordinates and likelihood scores are used to train a neural network. (5) A cut $\hat{s}_{\mathcal{L}, \rm min}$ in the predicted likelihood score is determined that corresponds to the likelihood score of the live set. (6) The new proposal volume is defined as that part of the bounding ellipsoid where the predicted likelihood score is above $\hat{s}_{\mathcal{L}, \rm min}$.}
    \label{fig:bounds}
\end{figure*}

The total sampling efficiency is maximal if, within each shell, the variance of likelihood values is minimal. Thus, the efficiency of the INS algorithms is maximal if the surfaces of the shells trace the iso-likelihood surfaces. In other words, we would like the new bounds to be very close to the volume of the prior space where $\mathcal{L} > \mathcal{L}_{\rm min}$.

Several approaches to estimating the iso-likelihood surface have been proposed in the literature and fall into two broad categories: region and step sampling. In region sampling, we directly sample uniformly from an analytically tractable volume. One popular approach is to draw multiple, possibly overlapping ellipsoids around the live set, which is the basis for \multinest{} \citep{Feroz2009_MNRAS_398_1601} and \dynesty{} \citep{Speagle2020_MNRAS_493_3132} with uniform sampling. Another approach implemented as the default in \ultranest{} is to sample from ellipsoids centred on each live point \citep{Buchner2016_SC_26_383}. In step sampling, we sample points by starting from and perturbing an existing random live point. As implemented in \dynesty{}, we may use a random walk with a fixed number of steps to perform the perturbation. Another approach, slice sampling, is the basis for {\sc PolyChord} \citep{Handley2015_MNRAS_453_4384} and is also available in \dynesty{} and \ultranest{}. Recall that for the NS algorithm to be unbiased, new points must be drawn uniformly from a volume fully encompassing the iso-likelihood surface. In practice, region samplers tend to violate this requirement by missing part of the space. Similarly, step samplers may violate this assumption by creating samples with non-negligile correlations with the live set.

\subsection{Neural networks}

Fortunately, sampling from a volume not fully encompassing the equi-likelihood surface will not necessarily bias the INS algorithm, only reduce its efficiency. However, we require estimates of the size of the proposal volumes. Thus, we choose a sampling strategy falling into the class of region samplers. However, a problem with existing region sampling strategies is that they only consider a small subset of the available information: the coordinates of points in the live set. Existing methods ignore information about the positions and likelihood values of points not in the live set as well as the likelihood values of points in the live set. Here, we present a new algorithm that uses all the available information obtained during the exploration phase. The basic steps of this algorithm are illustrated in Fig.~\ref{fig:bounds}.
\begin{enumerate}
    \item When constructing a new bound, first identify the live set among all points evaluated thus far.
    \item Draw an ellipsoid around the live set and identify all points that fall within that ellipsoid, whether in the live set or not. The ellipsoid is constructed by first drawing an approximate minimum-volume enclosing ellipsoid\footnote{\url{http://www.mathworks.com/matlabcentral/fileexchange/9542}} around the live set and then expanding it by an enlargement factor $\varepsilon$ in each dimension. \label{item:mvee}
    \item Use Cholesky decomposition to transform the initial coordinates $\mathbf{\Theta}$ of all points into the coordinate system of the ellipsoid. By construction, the transformed coordinates $\mathbf{\tilde{\Theta}}$ of all points in the ellipsoid obey $|\mathbf{\tilde{\Theta}}| \leq 1$. Similarly, normalise the likelihood values by assigning a likelihood score. For all points not in the live set, the score is $s_{\mathcal{L}} = 0.5 p_{\mathcal{L}}$, where $p_{\mathcal{L}}$ is the percentile of each point's likelihood among the points not in the live set. Similarly, for the live set, the score is defined as $s_{\mathcal{L}} = 0.5 (1 + p_{\mathcal{L}})$, with the percentile being defined with respect to points in the live set.
    \item Train an ensemble of $n_{\rm network}$ independent fully connected neural network regressors on the dependence of $s_{\mathcal{L}}$ on $\mathbf{\tilde{\Theta}}$. By taking the average of the prediction of the $n_{\rm network}$ networks, we can predict likelihood scores $\hat{s}_\mathcal{L}$ for arbitrary points in parameter space.
    \item Analyse the distribution of true versus mean estimated likelihood scores and determine the average estimated likelihood score $\hat{s}_{\mathcal{L}, \rm min}$ of points at the edge of the live set, i.e., those with $s_{\mathcal{L}} = 0.5$.
    \item Sample new points by sampling uniformly from the bounding ellipsoid and then only allowing points for which the predicted likelihood score $\hat{s}_{\mathcal{L}}$ is equal or larger than $\hat{s}_{\mathcal{L}, \rm min}$. \label{item:draw}
\end{enumerate}

When sampling new points in step \ref{item:draw}, the neural network will accept only a fraction $f_{\rm network}$ of points drawn from the bounding ellipsoid. An estimate for the volume of the bound is thus $f_{\rm network} V_{\rm ell}$, where $V_{\rm ell}$ is the volume of the ellipsoid from which points are drawn, which can be calculated analytically.

\subsection{Multi-ellipsoid decomposition}

The straightforward algorithm above can slow down substantially in high dimensions and for complex likelihoods when a single bounding ellipsoid would be much larger than the volume probed by live set. If that is the case, the neural network may accept only a negligible fraction of points in the ellipsoid. Additionally, the predictive performance of the networks might decrease since the likelihood scores would change on scales much smaller than order unity in $\tilde{\Theta}$. Thus, we also employ an ellipsoidal decomposition to pre-select interesting target volumes before refining that selection with neural network regression. Fortunately, we can be very conservative in constructing the multi-ellipsoidal decomposition since drawing points from ellipsoids and evaluating their likelihood score is extremely fast, i.e., of order $\mathcal{O} (\mu s)$ on average on a modern computer.

First, at step \ref{item:mvee}, we draw multiple ellipsoids around the live set if those ellipsoids do not overlap. This may be expected if the posterior shows multiple distinct modes. We then perform subsequent steps for each volume or mode separately, e.g., train multiple neural networks for each mode separately. Thus, every ensemble of networks only needs to characterize a single mode of the likelihood surface.

Similarly, at step \ref{item:draw}, we do not propose new points from the bounding ellipsoids used at step \ref{item:mvee}. Instead, we repeatedly split the largest ellipsoid into two smaller ellipsoids if the volume of the union of all ellipsoids is larger than $\beta \, \varepsilon^{N_{\rm dim}}$ times the volume of the live set and the two resulting ellipsoids have a volume smaller than the original one. Here, $\beta$ is a free parameter determining how aggressively ellipsoids are split and $\varepsilon$ is the ellipsoid enlargement factor mentioned earlier. By breaking the boundary into smaller and smaller ellipsoids, we ensure that the network accepts roughly one out of every $\beta \, \varepsilon^{N_{\rm dim}}$ points drawn randomly from the ellipsoid union. To perform the splitting, we need to know the volume of the live set, which one can estimate from the fraction of the live set in each bound. To draw new points uniformly from the ellipsoid union, we randomly select an ellipsoid with a probability proportional to its volume. If a newly drawn point is part of $n > 1$ ellipsoids, we reject the points with probability $1 - 1 / n$. Finally, we can use the fraction of rejected points due to ellipsoid overlap together with the individual ellipsoid volumes to estimate the volume of the ellipsoid union $V_{\rm ell}$ and, ultimately, the volume of the bound.

\section{Application}
\label{sec:application}

In this section, we test the neural network-boosted INS algorithm on various challenging problems, both synthetic likelihoods and real-world applications.

\subsection{Bayesian sampling codes}

We have implemented the proposed algorithm in an MIT-licensed, open-source code called \nautilus{}. In this work, we use \nautilus{} version $0.7$. By default, \nautilus{} uses $N_{\rm live} = 2000$, $N_{\rm update} = N_{\rm live}$, $\varepsilon = 1.1$ and $\beta = 100$. Furthermore, for the exploration phase, we require $f_{\rm live} < 0.01$ and for the sampling phase we require $N_{\rm eff} > 10,000$. Finally, we use the neural network regressor as implemented in the \texttt{MLPRegressor} class in {\sc scikit-learn} version $1.1.1$. For the network, we use three hidden layers with $100$, $50$, and $20$ neurons, each with ReLU activation functions and the Adam optimiser for training \citep{Kingma2014_arXiv_1412_6980}. We test \nautilus{} with (nautilus-r) and without (nautilus) discarding points drawn in the exploration phase, as described in section \ref{subsec:sampling_phase}.

We compare \nautilus{} against other widely-used codes, including the NS codes \dynesty{} \citep{Speagle2020_MNRAS_493_3132} version $2.0.3$ and \ultranest{} \citep{Buchner2016_SC_26_383} version $3.5.7$ as well as the Preconditioned Monte-Carlo \citep{Karamanis2022_MNRAS_516_1644} code \pocomc{} \citep{Karamanis2022_JOSS_7_4634} version $0.2.4$. For \dynesty{}, we run the dynamic sampler separately with uniform sampling (dynesty-u), random walk sampling (dynesty-r) and slice sampling along preferred orientations (dynesty-s). With \dynesty{} and \ultranest{} being NS codes, the hyper-parameters $N_{\rm live}$, $f_{\rm live}$ and $N_{\rm eff}$ can be defined in the same manner as for \nautilus{}. \dynesty{}, by default, uses $N_{\rm live} = 500$ and requires $f_{\rm live} < 0.01$ and $N_{\rm eff} > 10,000$, the same as \nautilus{}. For \ultranest{}, we use the dynamic sampler with the MLFriends region sampler (UltraNest-m). We note that \ultranest{} also supports step samplers, including several slice sampling modes, but we do not test this here. In principle, we would expect results \ultranest{} with slice sampling to be qualitatively similar to those of \dynesty{} with slice sampling. By default, \ultranest{} uses $N_{\rm live} = 400$ and requires $f_{\rm live} < 0.01$, $N_{\rm eff} > 400$ and an evidence uncertainty of $0.5$ or less in $\log \mathcal{Z}$.

\pocomc{}'s underlying algorithm, Preconditioned Monte-Carlo \citep{Karamanis2022_MNRAS_516_1644}, is based on combining Sequential Monte Carlo (SMC) with normalising flows. In essence, \pocomc{} evolves a sample of $N_{\rm p}$ particles from an initial distribution, typically the prior, to the posterior distribution via a series of intermediate distributions. Interestingly, this SMC approach shares strong similarities to NS \citep{Salomone2018_arXiv_1805_3924, Ashton2022_NRvMP_2_39}. When running \pocomc{}, we use $N_{\rm p} = 1000$. We refer the reader to \citep{Karamanis2022_MNRAS_516_1644} for a discussion of the other hyper-parameters of the algorithm. After the initial sampling, we add $9000$ particles for a total of $10,000$ particles and use the Gaussianised Bridge Sampling \citep{Jia2019_arXiv_1912_6073} sampling technique implemented in \pocomc{} to estimate the evidence. The posterior samples returned by \pocomc{} are equal-weighted and, ignoring correlations between samples, we approximate $N_{\rm eff} = 10,000$.

Unless otherwise noted, we run all samplers in their default configurations since tuning each sampler to each individual problem is beyond the scope of this work. In most cases, tuning the hyper-parameters of the samplers results in trade-offs between accuracy, sampling efficiency, and the total number of likelihood evaluations. For example, for NS algorithms with step sampling, i.e., \dynesty{} with random walk or slice sampling and {\sc PolyChord}, we may increase the number of steps (also called walks or repeats) before a new point is accepted into the live set. This is expected to increase accuracy at the cost of requiring more likelihood evaluations \citep[see][for an example in cosmology]{Lemos2023_MNRAS_521_1184}. Given all these considerations, in the following, one should simultaneously consider all aspects of the samplers, i.e., speed, accuracy, and efficiency, instead of focusing on single characteristics in isolation.

\subsection{Comparison metrics}

\begin{figure*}
    \centering
    \includegraphics[width=\textwidth]{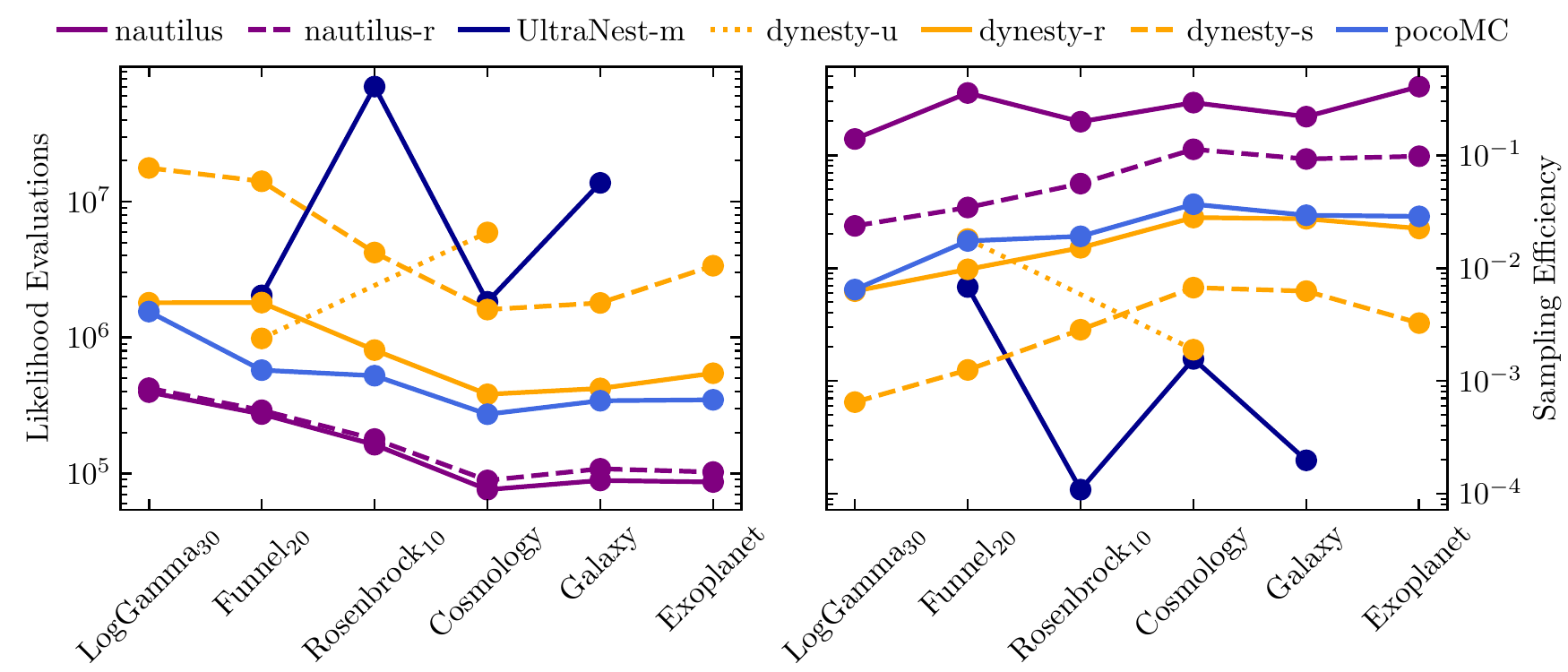}
    \caption{Number of likelihood evaluations (left) and sampling efficiency (right) for each sampler on a given problem. The sampling efficiency is defined as the effective sample size $N_{\rm eff}$ divided by the number of likelihood evaluations. All samplers were run in their default configurations.}
    \label{fig:performance}
\end{figure*}

\begin{table*}
    \begin{tabular}{lcccccc}
        Sampler & LogGamma$_{30}$ & Funnel$_{20}$ & Rosenbrock$_{10}$ & Cosmology & Galaxy & Exoplanet \\
        \hline
        analytic & $0$ & $0$ & -- & -- & -- & -- \\
        nautilus & $-0.035 \pm 0.029$ & $-0.100 \pm 0.003$ & $-43.145 \pm 0.023$ & $-24.663 \pm 0.007$ & $+395.783 \pm 0.007$ & $-98.899 \pm 0.006$ \\
nautilus-r & $-0.003 \pm 0.009$ & $-0.002 \pm 0.007$ & $-43.176 \pm 0.024$ & $-24.689 \pm 0.008$ & $+395.789 \pm 0.010$ & $-98.902 \pm 0.008$ \\
dynesty-u & -- & $+0.046 \pm 0.298$ & -- & $-24.680 \pm 0.165$ & -- & -- \\
dynesty-r & $+2.832 \pm 0.420$ & $+0.322 \pm 0.398$ & $-43.175 \pm 0.747$ & $-24.587 \pm 0.304$ & $+395.883 \pm 0.357$ & $-98.771 \pm 0.229$ \\
dynesty-s & $-0.021 \pm 0.333$ & $+0.049 \pm 0.331$ & $-43.108 \pm 0.551$ & $-24.647 \pm 0.198$ & $+395.857 \pm 0.263$ & $-98.902 \pm 0.163$ \\
pocoMC & $-1.001 \pm 0.210$ & $-1.179 \pm 1.015$ & $-43.745 \pm 0.222$ & $-24.739 \pm 0.017$ & $+395.707 \pm 0.021$ & $-99.029 \pm 0.040$ \\
UltraNest-m & -- & $+0.032 \pm 0.415$ & $-43.290 \pm 0.317$ & $-24.691 \pm 0.190$ & $+395.771 \pm 0.197$ & -- \\

    \end{tabular}
    \caption{Logarithm of the Bayesian evidence reported by different samplers for the problem discussed in this work. We report the mean and spread from repeated runs of each sampler on a given problem. We also state the analytically calculated evidence, if known.}
    \label{tab:evidence}
\end{table*}

\begin{table*}
    \begin{tabular}{lcccccc}
        Sampler & LogGamma$_{30}$ & Funnel$_{20}$ & Rosenbrock$_{10}$ & Cosmology & Galaxy & Exoplanet \\
        \hline
        nautilus & $33.49 \pm 1.08$ & $199.19 \pm 1.41$ & $18.01 \pm 0.21$ & $7.15 \pm 0.05$ & $5.76 \pm 0.04$ & $18.03 \pm 0.10$ \\
nautilus-r & $34.20 \pm 0.37$ & $198.09 \pm 2.58$ & $17.96 \pm 0.28$ & $7.14 \pm 0.08$ & $5.72 \pm 0.09$ & $17.88 \pm 0.18$ \\
dynesty-u & -- & $198.96 \pm 10.57$ & -- & $7.11 \pm 0.18$ & -- & -- \\
dynesty-r & $32.87 \pm 1.16$ & $198.89 \pm 14.42$ & $17.08 \pm 2.43$ & $6.84 \pm 0.32$ & $5.50 \pm 0.39$ & $17.21 \pm 0.95$ \\
dynesty-s & $34.23 \pm 1.22$ & $199.67 \pm 10.60$ & $17.50 \pm 2.30$ & $6.96 \pm 0.22$ & $5.65 \pm 0.27$ & $17.92 \pm 0.53$ \\
pocoMC & $26.59 \pm 1.05$ & $137.12 \pm 24.23$ & $16.31 \pm 0.83$ & $6.68 \pm 0.23$ & $5.17 \pm 0.21$ & $16.56 \pm 0.54$ \\
UltraNest-m & -- & $198.20 \pm 12.18$ & $18.43 \pm 0.91$ & $7.12 \pm 0.25$ & $5.79 \pm 0.23$ & -- \\

    \end{tabular}
    \caption{Same as Table~\protect\ref{tab:evidence} but for the BMD.}
    \label{tab:bmd}
\end{table*}

In the following, we will discuss the different test problems and results from the samplers. Two metrics we study are the total number of likelihood evaluations $N_{\rm like}$ and the total sampling efficiency $\eta$, defined as $N_{\rm eff} / N_{\rm like}$. Average numbers for these two metrics for all samplers and problems tested are shown in Fig.~\ref{fig:performance}. However, we note that our estimates for the effective sample size are approximate as eq.~\eqref{eq:ess} is only valid for independent samples. As we will see later, the effective sample size reported by \nautilus{} is fairly accurate, whereas the estimates for \dynesty{} with random walk sampling and \pocomc{} are likely too high. In addition to the total number of likelihood evaluations and sampling efficiency, Table \ref{tab:evidence} compares the different Bayesian evidence estimates for different problems regarding their means and scatters over repeated runs. Additionally, we compare the estimated Bayesian Model Dimensionality \citep[BMD, ][]{Handley2019_PhRvD_100_3512} $d$ in Table \ref{tab:bmd}, which can be defined as
\begin{equation}
    \frac{d}{2} = \langle \log \mathcal{L}^2 \rangle - \langle \log \mathcal{L} \rangle^2 \, ,
\end{equation}
where $\langle \rangle$ denotes the average over the posterior. This statistic can be straightforwardly estimated from the posterior distribution and is independent of estimates of the evidence. Typically, a too-low BMD estimate signals underestimated posterior uncertainties. Finally, we visually inspect the average one-dimensional posteriors. Unless stated otherwise, the reported results come from averaging $200$ repeated applications of the samplers to a given problem.

\subsection{LogGamma likelihood}

The so-called LogGamma likelihood has been identified as a challenging likelihood to sample from with NS algorithms \citep{Feroz2019_OJAp_2_10}. This likelihood, for which we use the definition given in \cite{Buchner2016_SC_26_383}, can be defined for arbitrary dimensionality $N_{\rm dim} \geq 2$. Along the first dimension, the likelihood is given by
\begin{equation}
    \mathcal{L}_1(\theta_1) = \frac{1}{2} \left[ f_{\log\Gamma} \left( \theta_1 \bigg\rvert 1, \frac{1}{3}, \frac{1}{30} \right) + f_{\log\Gamma} \left( \theta_1 \bigg\rvert 1, \frac{2}{3}, \frac{1}{30} \right) \right] \,
\end{equation}
where $f_{\log\Gamma} (x | \alpha, \mu, \sigma)$ denotes the value of the probability density function (PDF) of a log-gamma distribution with skew $\alpha$, location $\mu$ and scatter $\sigma$. Along the second dimension, the likelihood is
\begin{equation}
    \mathcal{L}_2(\theta_2) = \frac{1}{2} \left[ f_\mathcal{N} \left( \theta_2 \bigg\rvert \frac{1}{3}, \frac{1}{30} \right) + f_\mathcal{N} \left( \theta_2 \bigg\rvert \frac{2}{3}, \frac{1}{30} \right) \right] \, ,
\end{equation}
with $f_{\mathcal{N}} (x | \mu, \sigma)$ being the value of the PDF of a normal distribution with mean $\mu$ and scatter $\sigma$. For all other parameter dimensions $i$ from $3$ to less or equal $0.5 N_{\rm dim} + 1$, the likelihood is
\begin{equation}
    \mathcal{L}_i(\theta_i) = f_{\log\Gamma} \left( \theta_i \bigg\rvert 1, \frac{2}{3}, \frac{1}{30} \right)
\end{equation}
and for all other dimensions it is
\begin{equation}
    \mathcal{L}_i(\theta_i) = f_{\mathcal{N}} \left( \theta_i \bigg\rvert \frac{2}{3}, \frac{1}{30} \right) \, .
\end{equation}
With the final likelihood defined via
\begin{equation}
    \mathcal{L}(\mathbf{\Theta}) = 10^{N_{\rm dim}} \times \prod\limits_1^{N_{\rm dim}} \mathcal{L}_i(\theta_i) \, ,
\end{equation}
the Bayesian evidence $\mathcal{Z}$ can be calculated to be $\log \mathcal{Z} = 0$ if we assume a flat prior over $[-5, +5]$ in all dimensions. The difficulty of this likelihood stems from the use of multiple heavy-tailed log-gamma distributions. As indicated previously, ellipsoidal region samplers, such as the one employed by \multinest{}, tend to incorrectly exclude parts of the iso-likelihood surface at each iteration, leading to a too fast ``shrinking'' of the live set and an overall overestimated evidence \citep{Buchner2016_SC_26_383} while simultaneously under-estimating posterior uncertainties. More recent NS codes with region sampling, e.g., \ultranest{} with MLFriends and \dynesty{} with uniform sampling, have implemented leave-out cross-validation of the bounds to minimise this issue \citep{Buchner2016_SC_26_383, Speagle2020_MNRAS_493_3132}. While this noticeably increases accuracy, it also results in substantially larger numbers of likelihood evaluations, especially in higher dimensions \citep{Buchner2016_SC_26_383}. In fact, we were unable to test \dynesty{} with uniform sampling and \ultranest{} with MLFriends for $N_{\rm dim} = 30$ due to their excessively long runtimes \citep[also see][]{Buchner2016_SC_26_383}.

\begin{figure}
    \centering
    \includegraphics[width=\columnwidth]{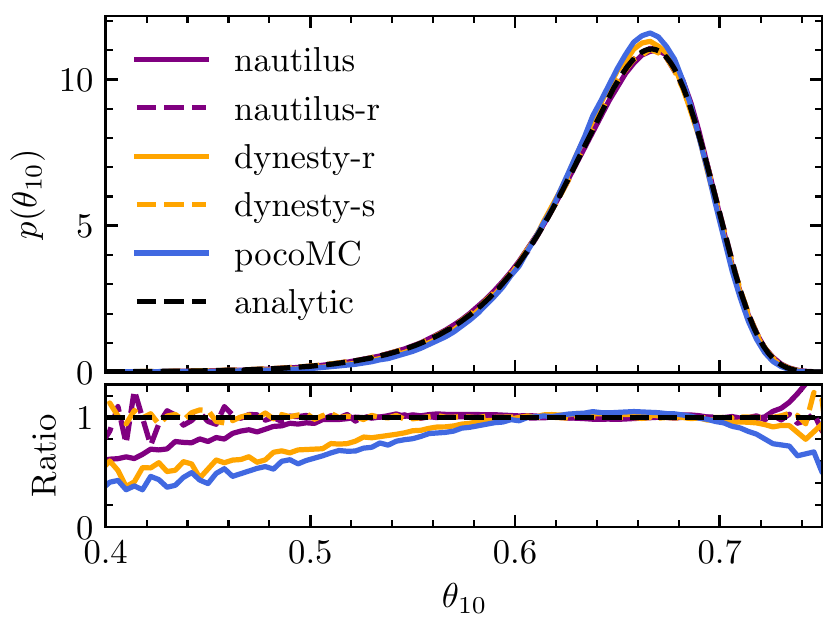}
    \caption{Estimated marginalised posterior distribution of the tenth parameter of the $30$-dimensional LogGamma distribution. In all cases, the results of individual samplers are averaged over repeated runs. We also overplot the analytic result. The bottom panel shows the ratio of the estimated marginal posterior distributions to the analytic result.}
    \label{fig:loggamma-30_x10_posterior}
\end{figure}

First, we test the LogGamma problem in $N_{\rm dim} = 30$ dimensions. When averaging the results over repeated runs, all samplers are able to recover one-dimensional posterior estimates that are close to the true posterior, which can be calculated analytically. In Fig.~\ref{fig:loggamma-30_x10_posterior} we show the posterior estimates of the tenth parameter, which is representative of the overall performance of the different samplers. We see that \nautilus{} after re-sampling as well as \dynesty{} with slice sampling are able to recover very accurate posterior estimates, even in the tails of the distribution. Without re-sampling, the posterior estimated with \nautilus{} is very slightly biased due to using a pseudo-importance function. The results from \pocomc{} and \dynesty{} with random walk sampling show much more significant biases. On average, for this likelihood problem, \pocomc{} and \dynesty{} with random walk sampling underestimate one-dimensional posterior uncertainties by around $6\%$ and $4\%$, respectively. Similarly, \pocomc{} produces substantially lower estimates of the BMD than other samplers, as shown in Table \ref{tab:bmd}. Looking at the evidence estimates in Table \ref{tab:evidence}, we see that \nautilus{} can recover the expected evidence to very high accuracy, especially after re-sampling the points in the initial exploration phase. In the latter case, the results are accurate to within at least sub-percent accuracy. \dynesty{} with slice sampling can also recover accurate evidence estimates albeit with $30$ times larger uncertainties. Contrary, \dynesty{} with random walk sampling and \pocomc{} are both unable to recover the proper evidence, clustering around $\log \mathcal{Z} \sim +3$ and $-1$, respectively.

Looking at the runtime and efficiency numbers in Fig.~\ref{fig:performance}, we see that \nautilus{} needs the lowest number of likelihood evaluations of all the samplers tested and also provides the highest sampling efficiency. The sampling efficiency of \nautilus{} run with re-sampling is nominally much lower than the default run. That is because we consider all likelihood estimates in the exploration and sampling phase when calculating the efficiency, even though points from the exploration phase are discarded. If we calculated the sampling efficiency taking only into account new likelihood evaluations during the sampling phase, the efficiency would be even higher. On average, after the initial exploration phase, \nautilus{} needs only roughly $40,000$ likelihood evaluations to increase the effective posterior sample size by $10,000$. Similar results regarding sampling efficiency after the exploration phase hold for the other problems in this section. As noted previously, our effective sample size estimates using eq.~\eqref{eq:ess} are not always fully accurate since the posterior samples are correlated. This is especially true for \dynesty{} with random walk sampling and \pocomc{} since they use step samplers. We verified that the effective sample sizes are indeed overestimates by looking at the scatter of the posterior estimates between different runs. We do not find this to be the case for \dynesty{} with slice sampling or for \nautilus{}.

\begin{figure}
    \centering
    \includegraphics{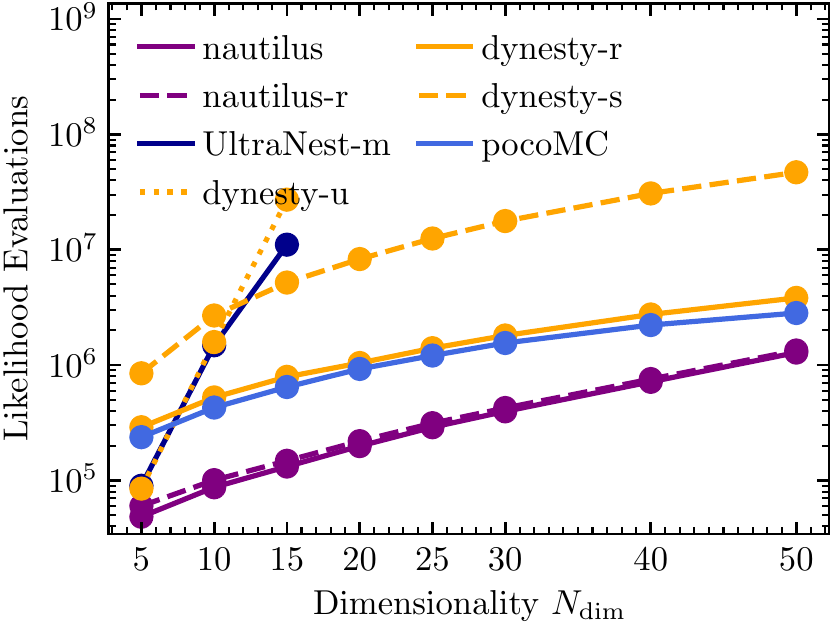}
    \caption{The dependence of the total number of likelihood evaluations on the dimensionality of the LogGamma problem.}
    \label{fig:scaling}
\end{figure}

Finally, we investigate the scaling of the different codes with the dimensionality of the problem up to $N_{\rm dim} = 50$ in Fig.~\ref{fig:scaling}. We see that \ultranest{} and \dynesty{} with uniform sampling, the two nested sampling implementations with region sampling, have a steep scaling with $N_{\rm dim}$. Codes based on step sampling, such as \dynesty{} with random walk and slice sampling, as well as \pocomc, have a more mild scaling. Interestingly, \nautilus{} has a scaling very similar to that of codes with step sampling despite being based on region sampling. While \nautilus{} scales slightly stronger with $N_{\rm dim}$ than \dynesty{} with random walk sampling and \pocomc, the latter two start producing significantly inaccurate evidence estimates around $N_{\rm dim} = 10$. For example, for $N_{\rm dim} = 50$, \dynesty{} with random walk sampling and \pocomc report $\log \mathcal{Z} = +11$ and $-5$, respectively, whereas \nautilus{} after re-sampling still reports $\log \mathcal{Z} \approx 0$. We find that beyond $N_{\rm dim} = 50$, \nautilus{} starts slowing down considerably, both in terms of computational overhead as well as sampling efficiency. Some of this may be overcome by adjusting hyper-parameters. However, we expect other samplers such as nested sampling with slice sampling or Hamiltonian Monte-Carlo to be more suitable for problems with a large number of parameters, i.e. $N_{\rm dim} \sim \mathcal{O}(100)$.

\subsection{Correlated funnel likelihood}

The correlated funnel likelihood, a variant of Neal's funnel \citep{Neal2003}, is given in \cite{Karamanis2020_arXiv_2002_6212} and can also be defined for arbitrary dimensions. In the first dimension, the likelihood follows a normal distribution with $0$ mean and unit variance,
\begin{equation}
    \mathcal{L}_1(\theta_1) = f_{\mathcal{N}} (\theta_1 | 0, 1) \, .
\end{equation}
The likelihood along all other dimensions is given by a multivariate normal distribution with $0$ mean and covariance $\Sigma(\theta_1)$
\begin{equation}
    \mathcal{L}_{>1}(\theta_2, \dots, \theta_{\rm N} | \theta_1) = f_{\mathcal{N}}(\theta_2, \dots, \theta_{\rm N} | 0, \Sigma(\theta_1) ) \, ,
\end{equation}
where
\begin{equation}
    \Sigma(\theta_1) = \exp(\theta_1) \times \begin{bmatrix} 
    1 & 0.95 & \dots & 0.95 & 0.95 \\
    0.95 & 1 & \dots & 0.95 & 0.95 \\
    \vdots & \vdots & \ddots & \vdots & \vdots \\
    0.95 & 0.95 & \dots & 1 & 0.95 \\
    0.95 & 0.95 &  \dots  & 0.95 & 1 
    \end{bmatrix} \, .
\end{equation}
With the likelihood defined as
\begin{equation}
    \mathcal{L}(\mathbf{\Theta}) = 20^{N_{\rm dim}} \times \mathcal{L}_1(\theta_1) \times \mathcal{L}_{>1}(\theta_2, \dots, \theta_{\rm N} | \theta_1) \, ,
\end{equation}
the Bayesian evidence $\mathcal{Z}$ can be calculated to be $\log \mathcal{Z} = 0$ if we assume a flat prior over the range $[-10, +10]$ for all dimensions. This likelihood problem is challenging because at $\theta_1 < 0$ the posterior covers a small volume with high likelihood and whereas it covers a large volume with low likelihood for $\theta_1 > 0$. This leads to large auto-correlation times for MCMC samplers \citep{Karamanis2020_arXiv_2002_6212}. Similarly, this problem presents a challenge for NS algorithms with region sampling since the proposal volume likely misses parts of the low-$\theta_1$ iso-likelihood surface due to the large volume differences along the first dimension. Here, we test the $20$-dimensional version of the correlated funnel distribution.

\begin{figure}
    \centering
    \includegraphics[width=\columnwidth]{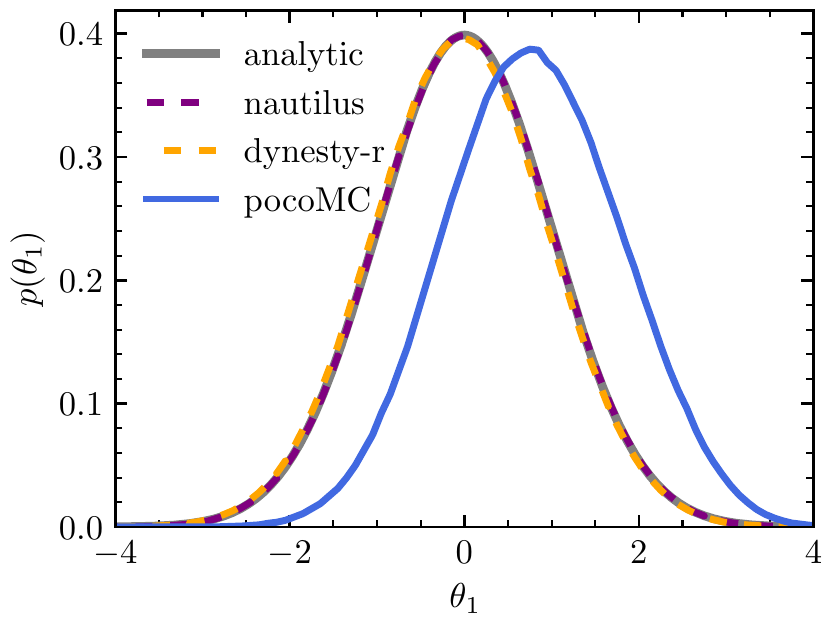}
    \caption{Average marginalised posterior estimates the first parameter of the $20$-dimensional correlated funnel distribution.}
    \label{fig:funnel-20_x1_posterior}
\end{figure}

We find that all samplers except \pocomc{} can recover very accurate posterior estimates. As an example, we show the mean marginalised posterior estimate for the first parameter in Fig.~\ref{fig:funnel-20_x1_posterior}. Similarly, as shown in Table \ref{tab:evidence}, all samplers except \pocomc{} produce very accurate Bayesian evidence estimates. As for the LogGamma likelihood, the evidence estimate produced by \nautilus{} is significantly more precise than that of other samplers by a factor of at least $30$. While not shown in Fig.~\ref{fig:funnel-20_x1_posterior}, the results from \pocomc{} can be made much more accurate by increasing the number of MCMC steps in each iteration at the cost of increasing the overall runtime. When comparing all samplers, \nautilus{} again needs the fewest likelihood evaluations by a factor of several and also has by far the highest sampling efficiency, similar to what was found for the LogGamma problem.

\subsection{Rosenbrock likelihood}

The Rosenbrock function is a popular test function for optimisation algorithms. We can define a corresponding likelihood function for arbitrary dimensions via
\begin{equation}
    \log \mathcal{L} (\mathbf{\Theta}) = \sum\limits_{i=1}^{N_{\rm dim} - 1} \left[ (1 - \theta_i)^2 + 100 \left( \theta_{i+1} - \theta_i^2 \right)^2 \right] \, .
\end{equation}
The difficulty of this likelihood problem lies in the strongly curved degeneracies between all parameters. Note that definition of the Rosenbrock likelihood in higher dimensions used here is more challenging compared to the one used in \cite{Karamanis2020_arXiv_2002_6212} and \cite{Jia2019_arXiv_1912_6073}, which represents $N_{\rm dim} / 2$ uncoupled two-dimensional Rosenbrock likelihoods. In this work, we test the $10$-dimensional Rosenbrock likelihood and employ a prior range of $[-5, +5]$. We chose not to run \dynesty{} with uniform sampling due to excessively long runtimes. Note that the likelihood definition and prior is the same as used in \cite{Moss2020_MNRAS_496_328} where \multinest{} and {\sc PolyChord} were tested along {\sc NNest}, an NS algorithm using normalising flows for proposals.

\begin{figure}
    \centering
    \includegraphics[width=\columnwidth]{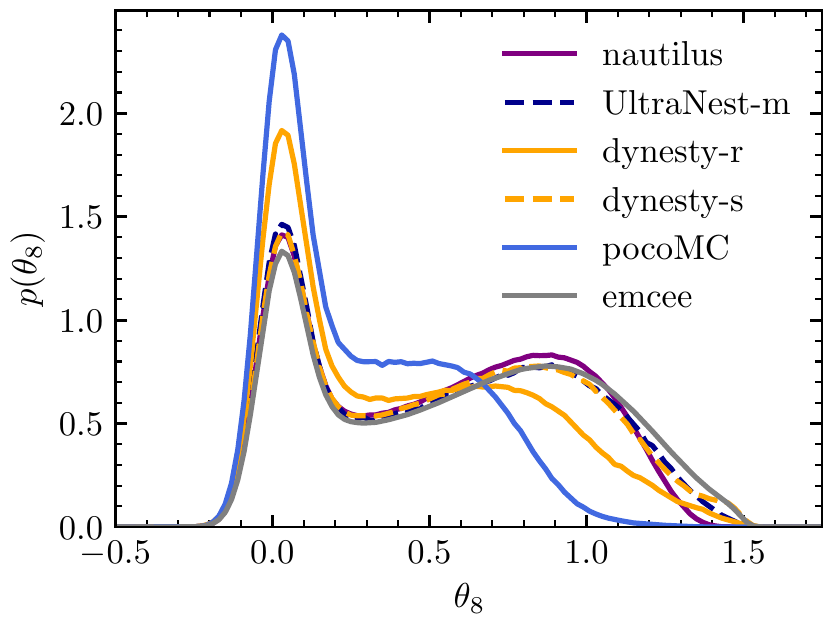}
    \caption{Average marginalised posterior estimates for the eighth parameter of the $10$-dimensional Rosenbrock likelihood.}
    \label{fig:rosenbrock-10_x8_posterior}
\end{figure}

\begin{figure*}
    \centering
    \includegraphics[width=\textwidth]{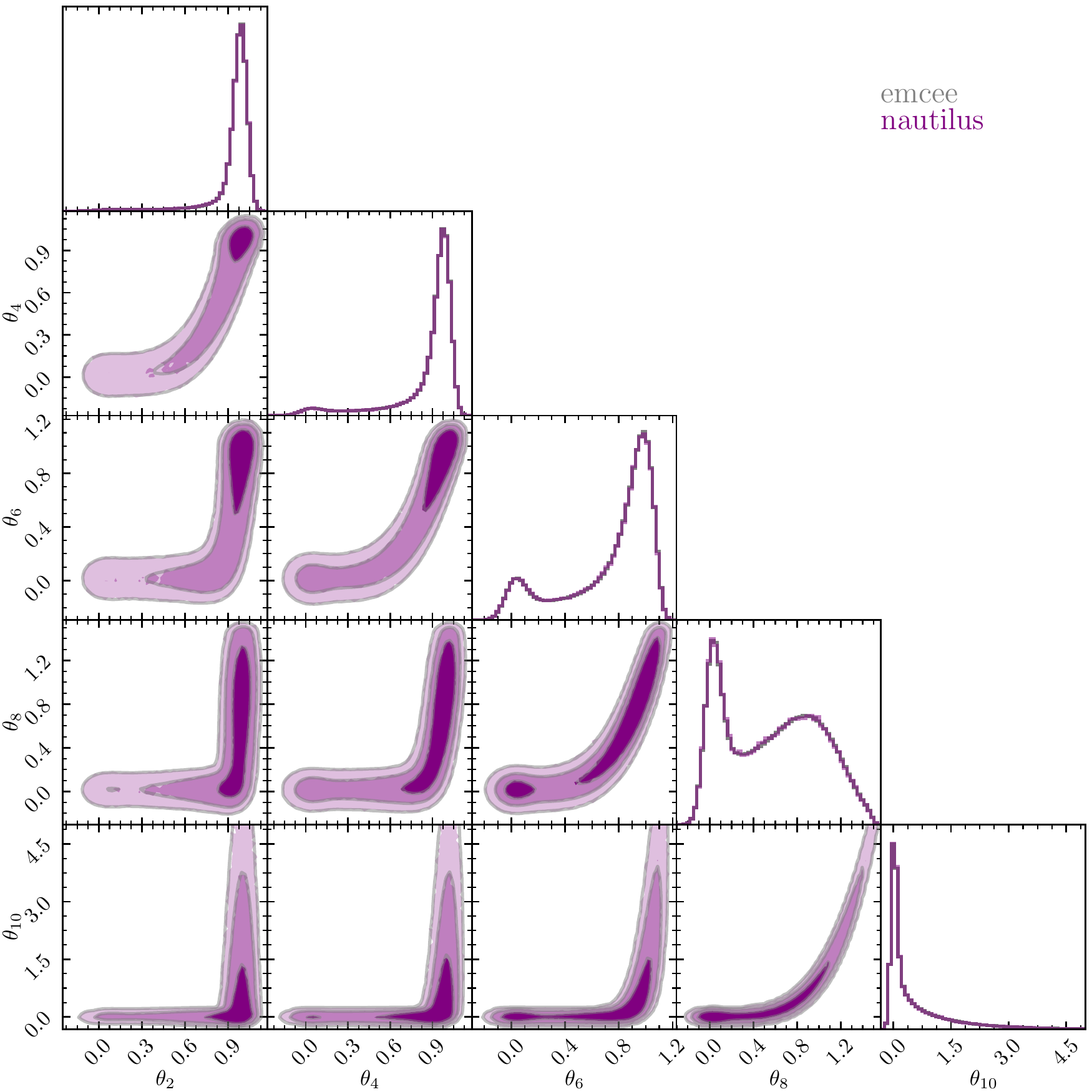}
    \caption{One and two-dimensional posteriors of the even-numbered parameters of the $10$-dimensional Rosenbrock likelihood. The results of the odd-numbered parameters are omitted for clarity and are qualitatively similar. The results of \emcee{} are shown by the grey contours and the estimates of \nautilus{} by the filling. We show the $68\%$, $95\%$ and $99.7\%$ containment ranges for the two-dimensional projections.}
    \label{fig:rosenbrock-10_full_posterior}
\end{figure*}

In Fig.~\ref{fig:rosenbrock-10_x8_posterior}, we show the one-dimensional posterior of the eighth parameter, as estimated by the different samplers, as an example. We find that all five samplers tested produce different results. We compare the results against the ones of the MCMC code \emcee{} \citep{ForemanMackey2013_PASP_125_306}, using $1000$ walkers and $10^8$ iterations, i.e., $10^{11}$ likelihood evaluations in total. This is necessary for a converged and accurate result since the auto-correlation time is of order $10^6$ steps. \nautilus{} produces results that come very close to the true posterior, whereas \dynesty{} with random walk sampling and \pocomc{} show more substantial biases. While the results of \nautilus{} are biased in the default configuration, the posteriors converge quickly when the number of live points is increased. In Fig.~\ref{fig:rosenbrock-10_full_posterior}, we show the one and two-dimensional posteriors of all even parameters. We compare the results of \emcee{} with the estimates of \nautilus{} run with $10,000$ live points in Fig.~\ref{fig:rosenbrock-10_full_posterior} and find that they are fully consistent, providing further evidence that both are producing accurate results. In this case, \nautilus{} needs around $8 \times 10^5$ likelihood evaluations, several orders of magnitude fewer than \emcee{}. While \pocomc{} and \dynesty{} with random walk sampling need similar numbers of likelihood evaluations, their results are almost certainly very inaccurate.

Finally, \cite{Moss2020_MNRAS_496_328} find that \multinest{}, {\sc PolyChord} and {\sc NNest} need roughly $3.9 \times 10^5$, $9.7 \times 10^6$ and $1.5 \times 10^6$ likelihood evaluations for the same problem, respectively. In particular, \multinest{} needs very few likelihood evaluations, of the same order as \nautilus{}. However, we cannot verify here whether any of the results obtained in \cite{Moss2020_MNRAS_496_328} were unbiased. Given that \multinest{} estimates $\log \mathcal{Z} \approx -42.1$ whereas {\sc PolyChord}, \ultranest{} with MLFriends, \dynesty{} with slice sampling, {\sc NNest} and \nautilus{} all favour $\sim -43.2$, it seems unlikely that the \multinest{} posterior and evidence estimate reported in \cite{Moss2020_MNRAS_496_328} is accurate.

\subsection{Cosmology application}

As an example of a real-world application to cosmology, we re-perform the analysis reported in \cite{Zentner2019_MNRAS_485_1196} by fitting parameters of the so-called galaxy--halo connection \citep{Hearin2017_AJ_154_190} to galaxy clustering data. Specifically, we are fitting the $M_r < -20$ sample using updated clustering measurements from \cite{Guo2015_MNRAS_453_4368} with an assembly bias model. This likelihood problem has seven dimensions. We refer the reader to \cite{Zentner2019_MNRAS_485_1196}, particularly Fig.~6, for a presentation of the posterior. To speed up the calculation, we use {\sc TabCorr} \citep{Lange2019_MNRAS_490_1870} version $1.0.0$ to make fast and accurate galaxy clustering model predictions \citep{Zheng2016_MNRAS_458_4015}. In their original analysis, \cite{Zentner2019_MNRAS_485_1196} used the \emcee{} package for posterior analysis and reported needing around $3 - 10 \times 10^6$ likelihood evaluations for converged results. This problem represents a less challenging example than the synthetic likelihood discussed above. Consequently, all samplers provide very similar and likely highly accurate posterior and evidence estimates. Among all samplers, \nautilus{} needs the fewest likelihood evaluations by a factor of several, requiring only $\sim 80,000$ likelihood estimations, two orders of magnitude fewer than in \cite{Zentner2019_MNRAS_485_1196}. Similarly, \nautilus{} has the highest sampling efficiency by an order of magnitude and produces the most precise evidence estimate together with \pocomc{}.

\subsection{Galaxy application}

The observed broad-band photometry or spectral energy distribution (SED) of a galaxy can be used to estimate its physical properties, such as redshift, stellar mass or star formation rate. At the same time, the observed galaxy SED often allows for strong, sometimes multi-modal degeneracies between different model parameters and can exhibit strong dependencies on assumed priors and model assumptions \citep{Pacifici2023_ApJ_944_141}. Therefore, rigorously quantifying posterior uncertainties is critical when studying the properties of large populations of observed galaxies. Unfortunately, the computational cost of fully Bayesian error quantification can be quite high, given the number of galaxies observed in modern galaxy surveys. For example, \cite{Leja2019_ApJ_877_140} used 1.5 million CPU hours to quantify the properties of just $\sim 60,000$ galaxies. While advanced machine learning-based approaches exist to overcome this problem \citep{Alsing2020_ApJS_249_5, Hahn2022_ApJ_938_11}, those require domain knowledge and often a large computational upfront cost, making them primarily suitable for analysing extremely large galaxy populations. Thus, more efficient Bayesian sampling would make fully Bayesian SED analyses more widely applicable. In this work, we test fitting the SED of a random galaxy in the CANDELS GOODS-South catalog \citep{Guo2013_ApJS_207_24} with the {\sc} spectral and SED fitting code {\sc bagpipes} \citep{Carnall2018_MNRAS_480_4379} version $1.0.0$. In this example, the model has seven free parameters. Due to the long runtimes, we did not run \dynesty{} with uniform sampling, and only $40$ runs with \ultranest{}. All the samplers tested produce very similar results regarding the posterior and evidence. Among all samplers tested, \nautilus{} has the shortest runtime, needing around $100,000$ likelihood evaluations, and the highest sampling efficiency.

\subsection{Exoplanet application}

A star's observed radial velocity (RV) can be used to detect the presence of one or several exoplanets orbiting it. In addition to computing the posterior distribution of exoplanet parameters such as mass, we might be particularly interested in estimating the Bayesian evidence. For example, by comparing the evidence ratio of a model with $n$ to a model with $n + 1$ exoplanets, one can quantify the detection significance of an additional exoplanet \citep{Nelson2020_AJ_159_73}. Here, we test the ability of different algorithms to estimate posteriors and Bayesian evidences when analysing the RV curve of K2-24 with a model for two planets, K2-24b and K2-24c \citep{Petigura2016_ApJ_818_36}. This problem, which has $14$ free parameters, is also described in the documentation of the {\sc exoplanet} package version $0.5.4$ \citep{ForemanMackey2021_JOSS_6_3285}. We did not run \dynesty{} with uniform sampling and \ultranest{} on this problem due to the long runtime.

We find that all samplers produce very similar posterior distributions. As in the previous examples, \pocomc{} and \dynesty{} with random walk sampling tend to report lower parameter uncertainties by $1$ to $2\%$, which can also be seen in the lower estimates of the BMD in Table~\ref{tab:bmd}. All samplers report very similar Bayesian evidence estimates, agreeing to within $\Delta \log \mathcal{Z} = 0.2$ when comparing averages over repeated runs. Finally, \nautilus{} needs the fewest likelihood evaluations, fewer than $100,000$ on average, to arrive at a result, and also has the highest sampling efficiency.

\section{Discussion}
\label{sec:discussion}

The results in the previous section demonstrate the excellent performance of the neural network-boosted INS algorithm as implemented in \nautilus{}. For all problems considered, by default, \nautilus{} always needs fewer likelihood evaluations and has a substantially higher sampling efficiency than all other samplers tested, including \dynesty{}, \ultranest{} and \pocomc{}. In fact, after the initial exploration phase, \nautilus{} has often constructed an importance function so close to the posterior that the latter can be sampled with $25\%$ to near $100\%$ efficiency. At the same time, the posterior and evidence estimates are highly accurate and/or agree well with that of other robust samplers such us \dynesty{} with slice sampling. In particular, we do not observe any apparent bias in the evidence estimates, unlike for the INS implementation in \multinest{} \citep[see, e.g.][]{Lemos2023_MNRAS_521_1184}. Furthermore, we do not observe any strong biases of the results due to using a pseudo-importance function. In any case, we also show how to remove that residual bias by re-sampling points obtained during the exploration phase in case of percent accuracy on the posterior or evidence being required. Besides the good sampling performance in the problems tested here, the boosted INS algorithm is also embarrassingly parallelisable by distributing likelihood evaluations over multiple CPUs. It should scale well up to $\sim N_{\rm update}$, i.e., thousands of CPUs.

\subsection{Sampling performance}

Many popular Bayesian sampling codes based on MCMC, NS and PMC have two drawbacks. First, only a fraction of calculated likelihood values is used to directly estimate posterior and evidence, while most proposals are commonly rejected. Second, only a fraction of the likelihood computations is used to inform new proposals. For example, MCMC algorithms typically base new proposals only on the current state of the chain, not past iterations. Similarly, NS implementations typically only use the positions of points in the live set to construct new samples but ignore rejected and inactive points and any detailed information on the likelihood of individual points. The INS algorithm presented in \cite{Feroz2019_OJAp_2_10} naturally overcomes the first issue, allowing us to use all likelihood evaluations to inform posterior and Bayesian evidences directly. In this work, in addition to presenting an improved INS algorithm, we show in section \ref{sec:bounds} how we can further boost the INS algorithm's efficiency with deep learning. In particular, this allows us to address the second issue: new proposals are informed by all previous likelihood evaluations. We believe these two key differences qualitatively explain the excellent properties of the neural network-boosted INS algorithm. We note that during the preparation of this manuscript, \cite{Williams2023_arXiv_2302_8526} introduced a different INS implementation based on deep learning with normalising flows, {\sc i-nessai}. While {\sc i-nessai} can also use all proposals for posterior and evidence estimation, it only uses the positions of points in the live set to make new proposals. There are also other differences between {\sc i-nessai} and \nautilus{} such as the construction of the importance function. We leave a detailed comparison to future work.

\subsection{Computational overhead}

The algorithm proposed here and implemented in \nautilus{} requires training several neural networks on the fly. However, the computational overhead introduced by this is limited. For example, for the real-world applications in this work, the time spent on training the networks is of the order of a few minutes on a Ryzen 27000U laptop CPU from 2018. In fact, even for likelihoods with virtually no computational costs, such as the synthetic likelihoods in section \ref{sec:application}, \nautilus{} typically needed roughly the same wall time as, for example, \dynesty{} or \ultranest{}. The overhead may be further reduced by, for example,  implementing neural networks on GPUs or more efficient neural network implementations than the one used in {\sc scikit-learn}. Similarly, Moore's law will increase the computational power available to Bayesian studies over time. While this allows for the evaluation of computationally more expensive likelihoods, the overhead of the boosted INS algorithm remains approximately constant and, over time, will constitute a smaller and smaller fraction of the total runtime. Overall, we do not expect neural network training to represent a significant bottleneck when running the boosted INS algorithm. Nonetheless, we see \nautilus{} as particularly well suited for expensive likelihood and if high accuracy or precision, i.e., large $N_{\rm eff}$, is required. For computationally inexpensive likelihoods where one only needs an approximate answer, other low-overhead, potentially more approximate algorithms such as \multinest{} with a small number of live points and high target efficiency may get results with lower computational costs \citep[see, e.g.,][section 7.3]{Feroz2009_MNRAS_398_1601}.

\subsection{Dimensionality scaling}

In Fig.~\ref{fig:scaling}, we explored how \nautilus{} scales with dimensionality, finding that it behaves similarly to region slice samplers. However, we note that the actual scaling will also depend on the difficulty of the likelihood surface, i.e., how accurately the neural networks can approximate it. Thus, for certain low-dimensional problems, \nautilus{} may have comparatively large runtime. In any case, parts of the algorithm presented here will not perform well for a large $N_{\rm dim}$. As described in section \ref{sec:bounds}, to propose new points, we first draw points from a union of ellipsoids and then only consider those that the neural network estimates to have a high likelihood. For large $N_{\rm dim}$, the ratio between the volume of the ellipsoids and the volume of the parameter space where the network predicts a high likelihood may become negligibly small. In this case, a large fraction of the wall time would be spent on sampling and rejecting points from the ellipsoid union without evaluating likelihoods. For this reason, we do not expect the current version of \nautilus{} to work well for $N_{\rm dim} \gtrsim 50$. In the future, one may use normalising flows instead of multi-dimensional ellipsoids to generate new proposals for the neural network regressor, potentially eliminating this bottleneck. We leave such studies to future work.

\subsection{Convergence and accuracy}

Finally, we want to stress the need for convergence studies like for any other Bayesian sampler. As shown in the Rosenbrock example, the algorithm can produce inaccurate results under certain circumstances. In the case of the Rosenbrock function, part of the high-likelihood region was missed during the exploration phase. In theory, all parts of the parameter space have a non-zero sampling probability such that importance weighting is expected to give an accurate result. In practice, part of the high-likelihood region may be assigned such a low sampling probability that they are effectively never sampled. We suggest that scientific studies using \nautilus{} should ensure that results are robust with respect to changing hyper-parameters, particularly the number of live points. Additionally, we recommend discarding points drawn during the exploration phase if using the results for scientific publications. As shown in section~\ref{sec:application}, not doing so may result in percent-level biases on the posterior and evidence estimates.

\section{Conclusion}
\label{sec:conclusion}

Given the prevalence of Bayesian inference in astronomy and cosmology and its computational cost, there is great demand for efficient and accurate Bayesian sampling algorithms. In this work, we have presented an updated version of the importance nested sampling \citep[INS, ][]{Feroz2019_OJAp_2_10} algorithm. Compared to the initial INS algorithm implemented in {\sc MultiNest}, our new algorithm allows for additional high-efficiency sampling after the initial algorithm is finished, akin to dynamic nested sampling \citep{Higson2019_SC_29_891} as implemented in, e.g., \dynesty{}. Furthermore, we introduced a new method to utilise deep learning to boost the sampling efficiency further. This allows the boosted INS algorithm to use the information about all previously sampled points to decide where new points are proposed. Another strength of this algorithm is that it is embarrassingly parallel and should scale well up to hundreds of CPUs. We also introduce an open-source, MIT-licensed implementation of the algorithm called \nautilus{}.

We tested \nautilus{} against a variety of established samplers, including \dynesty{}, \ultranest, and \pocomc{}, in synthetic problems and real-world applications. We find that in all example applications, \nautilus{} needs by far the fewest likelihood evaluations among all samplers tested. Similarly, it always has the highest sampling efficiency, returning the largest effective posterior sample size for a given number of likelihood evaluations. Likewise, \nautilus{} delivers by far the most precise evidence estimates with a relative scatter of only $\sim 1\%$. Most importantly, it achieves all this while returning highly accurate results for the benchmark problems. Comparing \nautilus{} only against samplers of similar accuracy, such as \dynesty{} with slice sampling, the reduction in the total number of likelihood evaluations is even greater. Given the results in our example applications, we believe the boosted INS algorithm as implemented in \nautilus{} and available at \url{https://github.com/johannesulf/nautilus} can be a valuable resource for researchers performing Bayesian inference studies.

\section*{Acknowledgements}

We thank Noah Weaverdyck, Josh Speagle, Johannes Buchner, and Minas Karamanis for helpful discussions. We are grateful to Lisalotte Crampton for designing the logo for \nautilus{}. We acknowledge use of the the lux supercomputer at UC Santa Cruz, funded by NSF MRI grant AST 1828315. We acknowledge support from the Leinweber Center for Theoretical Physics, NASA grant under contract 19-ATP19-0058 and DOE under contract DE-FG02-95ER40899.

This work made use of the following software packages: {\sc Astropy} \citep{AstropyCollaboration2013_AA_558_33}, {\sc bagpipes} \citep{Carnall2018_MNRAS_480_4379}, {\sc bibmanager} \citep{Cubillos2020_zndo_25_7042}, \dynesty{} \citep{Speagle2020_MNRAS_493_3132}, \emcee{} \citep{ForemanMackey2013_PASP_125_306}, {\sc halotools} \citep{Hearin2017_AJ_154_190}, {\sc NumPy} \citep{vanderWalt2011_CSE_13_22}, {\sc matplotlib} \citep{Hunter2007_CSE_9_90}, \pocomc{} \citep{Karamanis2022_JOSS_7_4634} {\sc scikit-learn} \citep{Pedregosa2011_JMLR_12_2825}, {\sc SciPy}, {\sc Spyder}, {\sc Setzer}, {\sc TabCorr} \citep{Lange2019_MNRAS_490_1870}, and \ultranest{} \citep{Buchner2016_SC_26_383}.

\section*{Data Availability}

All the code used to produce the results in this paper is available in the {\sc paper} branch of the \nautilus{} repository at \url{https://github.com/johannesulf/nautilus/tree/paper}. The results of all computations is available upon reasonable request to the author.

\bibliographystyle{mnras}
\bibliography{bibliography}

\begin{thebibliography}{}
\makeatletter
\relax
\def\mn@urlcharsother{\let\do\@makeother \do\$\do\&\do\#\do\^\do\_\do\%\do\~}
\def\mn@doi{\begingroup\mn@urlcharsother \@ifnextchar [ {\mn@doi@}
  {\mn@doi@[]}}
\def\mn@doi@[#1]#2{\def\@tempa{#1}\ifx\@tempa\@empty \href
  {http://dx.doi.org/#2} {doi:#2}\else \href {http://dx.doi.org/#2} {#1}\fi
  \endgroup}
\def\mn@eprint#1#2{\mn@eprint@#1:#2::\@nil}
\def\mn@eprint@arXiv#1{\href {http://arxiv.org/abs/#1} {{\tt arXiv:#1}}}
\def\mn@eprint@dblp#1{\href {http://dblp.uni-trier.de/rec/bibtex/#1.xml}
  {dblp:#1}}
\def\mn@eprint@#1:#2:#3:#4\@nil{\def\@tempa {#1}\def\@tempb {#2}\def\@tempc
  {#3}\ifx \@tempc \@empty \let \@tempc \@tempb \let \@tempb \@tempa \fi \ifx
  \@tempb \@empty \def\@tempb {arXiv}\fi \@ifundefined
  {mn@eprint@\@tempb}{\@tempb:\@tempc}{\expandafter \expandafter \csname
  mn@eprint@\@tempb\endcsname \expandafter{\@tempc}}}

\bibitem[\protect\citeauthoryear{{Alsing} \& {Handley}}{{Alsing} \&
  {Handley}}{2021}]{Alsing2021_MNRAS_505_95}
{Alsing} J.,  {Handley} W.,  2021, \mn@doi [\mnras] {10.1093/mnrasl/slab057},
  \href {https://ui.adsabs.harvard.edu/abs/2021MNRAS.505L..95A} {505, L95}

\bibitem[\protect\citeauthoryear{{Alsing} et~al.,}{{Alsing}
  et~al.}{2020}]{Alsing2020_ApJS_249_5}
{Alsing} J.,  et~al., 2020, \mn@doi [\apjs] {10.3847/1538-4365/ab917f}, \href
  {https://ui.adsabs.harvard.edu/abs/2020ApJS..249....5A} {249, 5}

\bibitem[\protect\citeauthoryear{{Ashton} et~al.,}{{Ashton}
  et~al.}{2022}]{Ashton2022_NRvMP_2_39}
{Ashton} G.,  et~al., 2022, \mn@doi [Nature Reviews Methods Primers]
  {10.1038/s43586-022-00121-x}, \href
  {https://ui.adsabs.harvard.edu/abs/2022NRvMP...2...39A} {2, 39}

\bibitem[\protect\citeauthoryear{{Astropy Collaboration} et~al.,}{{Astropy
  Collaboration} et~al.}{2013}]{AstropyCollaboration2013_AA_558_33}
{Astropy Collaboration} et~al., 2013, \mn@doi [\aap]
  {10.1051/0004-6361/201322068}, \href
  {https://ui.adsabs.harvard.edu/abs/2013A&A...558A..33A} {558, A33}

\bibitem[\protect\citeauthoryear{{Buchner}}{{Buchner}}{2016}]{Buchner2016_SC_26_383}
{Buchner} J.,  2016, \mn@doi [Statistics and Computing]
  {10.1007/s11222-014-9512-y}, \href
  {https://ui.adsabs.harvard.edu/abs/2016S&C....26..383B} {26, 383}

\bibitem[\protect\citeauthoryear{{Carnall}, {McLure}, {Dunlop}  \&
  {Dav{\'e}}}{{Carnall} et~al.}{2018}]{Carnall2018_MNRAS_480_4379}
{Carnall} A.~C.,  {McLure} R.~J.,  {Dunlop} J.~S.,   {Dav{\'e}} R.,  2018,
  \mn@doi [\mnras] {10.1093/mnras/sty2169}, \href
  {https://ui.adsabs.harvard.edu/abs/2018MNRAS.480.4379C} {480, 4379}

\bibitem[\protect\citeauthoryear{{Cubillos}}{{Cubillos}}{2020}]{Cubillos2020_zndo_25_7042}
{Cubillos} P.~E.,  2020, {bibmanager: A BibTeX manager for LaTeX projects},
  Zenodo, \mn@doi{10.5281/zenodo.2547042}

\bibitem[\protect\citeauthoryear{{Feroz}, {Hobson}  \& {Bridges}}{{Feroz}
  et~al.}{2009}]{Feroz2009_MNRAS_398_1601}
{Feroz} F.,  {Hobson} M.~P.,   {Bridges} M.,  2009, \mn@doi [\mnras]
  {10.1111/j.1365-2966.2009.14548.x}, \href
  {https://ui.adsabs.harvard.edu/abs/2009MNRAS.398.1601F} {398, 1601}

\bibitem[\protect\citeauthoryear{{Feroz}, {Hobson}, {Cameron}  \&
  {Pettitt}}{{Feroz} et~al.}{2019}]{Feroz2019_OJAp_2_10}
{Feroz} F.,  {Hobson} M.~P.,  {Cameron} E.,   {Pettitt} A.~N.,  2019, \mn@doi
  [The Open Journal of Astrophysics] {10.21105/astro.1306.2144}, \href
  {https://ui.adsabs.harvard.edu/abs/2019OJAp....2E..10F} {2, 10}

\bibitem[\protect\citeauthoryear{{Foreman-Mackey}, {Hogg}, {Lang}  \&
  {Goodman}}{{Foreman-Mackey} et~al.}{2013}]{ForemanMackey2013_PASP_125_306}
{Foreman-Mackey} D.,  {Hogg} D.~W.,  {Lang} D.,   {Goodman} J.,  2013, \mn@doi
  [\pasp] {10.1086/670067}, \href
  {https://ui.adsabs.harvard.edu/abs/2013PASP..125..306F} {125, 306}

\bibitem[\protect\citeauthoryear{{Foreman-Mackey} et~al.,}{{Foreman-Mackey}
  et~al.}{2021}]{ForemanMackey2021_JOSS_6_3285}
{Foreman-Mackey} D.,  et~al., 2021, \mn@doi [The Journal of Open Source
  Software] {10.21105/joss.03285}, \href
  {https://ui.adsabs.harvard.edu/abs/2021JOSS....6.3285F} {6, 3285}

\bibitem[\protect\citeauthoryear{{Guo} et~al.,}{{Guo}
  et~al.}{2013}]{Guo2013_ApJS_207_24}
{Guo} Y.,  et~al., 2013, \mn@doi [\apjs] {10.1088/0067-0049/207/2/24}, \href
  {https://ui.adsabs.harvard.edu/abs/2013ApJS..207...24G} {207, 24}

\bibitem[\protect\citeauthoryear{{Guo} et~al.,}{{Guo}
  et~al.}{2015}]{Guo2015_MNRAS_453_4368}
{Guo} H.,  et~al., 2015, \mn@doi [\mnras] {10.1093/mnras/stv1966}, \href
  {https://ui.adsabs.harvard.edu/abs/2015MNRAS.453.4368G} {453, 4368}

\bibitem[\protect\citeauthoryear{{Hahn} \& {Melchior}}{{Hahn} \&
  {Melchior}}{2022}]{Hahn2022_ApJ_938_11}
{Hahn} C.,  {Melchior} P.,  2022, \mn@doi [\apj] {10.3847/1538-4357/ac7b84},
  \href {https://ui.adsabs.harvard.edu/abs/2022ApJ...938...11H} {938, 11}

\bibitem[\protect\citeauthoryear{{Handley} \& {Lemos}}{{Handley} \&
  {Lemos}}{2019}]{Handley2019_PhRvD_100_3512}
{Handley} W.,  {Lemos} P.,  2019, \mn@doi [\prd] {10.1103/PhysRevD.100.023512},
  \href {https://ui.adsabs.harvard.edu/abs/2019PhRvD.100b3512H} {100, 023512}

\bibitem[\protect\citeauthoryear{{Handley}, {Hobson}  \& {Lasenby}}{{Handley}
  et~al.}{2015}]{Handley2015_MNRAS_453_4384}
{Handley} W.~J.,  {Hobson} M.~P.,   {Lasenby} A.~N.,  2015, \mn@doi [\mnras]
  {10.1093/mnras/stv1911}, \href
  {https://ui.adsabs.harvard.edu/abs/2015MNRAS.453.4384H} {453, 4384}

\bibitem[\protect\citeauthoryear{{Hearin} et~al.,}{{Hearin}
  et~al.}{2017}]{Hearin2017_AJ_154_190}
{Hearin} A.~P.,  et~al., 2017, \mn@doi [\aj] {10.3847/1538-3881/aa859f}, \href
  {https://ui.adsabs.harvard.edu/abs/2017AJ....154..190H} {154, 190}

\bibitem[\protect\citeauthoryear{{Higson}, {Handley}, {Hobson}  \&
  {Lasenby}}{{Higson} et~al.}{2019}]{Higson2019_SC_29_891}
{Higson} E.,  {Handley} W.,  {Hobson} M.,   {Lasenby} A.,  2019, \mn@doi
  [Statistics and Computing] {10.1007/s11222-018-9844-0}, \href
  {https://ui.adsabs.harvard.edu/abs/2019S&C....29..891H} {29, 891}

\bibitem[\protect\citeauthoryear{{Hunter}}{{Hunter}}{2007}]{Hunter2007_CSE_9_90}
{Hunter} J.~D.,  2007, \mn@doi [Computing in Science and Engineering]
  {10.1109/MCSE.2007.55}, \href
  {https://ui.adsabs.harvard.edu/abs/2007CSE.....9...90H} {9, 90}

\bibitem[\protect\citeauthoryear{{Jia} \& {Seljak}}{{Jia} \&
  {Seljak}}{2019}]{Jia2019_arXiv_1912_6073}
{Jia} H.,  {Seljak} U.,  2019, \mn@doi [arXiv e-prints]
  {10.48550/arXiv.1912.06073}, \href
  {https://ui.adsabs.harvard.edu/abs/2019arXiv191206073J} {p. arXiv:1912.06073}

\bibitem[\protect\citeauthoryear{{Karamanis} \& {Beutler}}{{Karamanis} \&
  {Beutler}}{2020}]{Karamanis2020_arXiv_2002_6212}
{Karamanis} M.,  {Beutler} F.,  2020, \mn@doi [arXiv e-prints]
  {10.48550/arXiv.2002.06212}, \href
  {https://ui.adsabs.harvard.edu/abs/2020arXiv200206212K} {p. arXiv:2002.06212}

\bibitem[\protect\citeauthoryear{{Karamanis}, {Nabergoj}, {Beutler}, {Peacock}
  \& {Seljak}}{{Karamanis} et~al.}{2022a}]{Karamanis2022_JOSS_7_4634}
{Karamanis} M.,  {Nabergoj} D.,  {Beutler} F.,  {Peacock} J.,   {Seljak} U.,
  2022a, \mn@doi [The Journal of Open Source Software] {10.21105/joss.04634},
  \href {https://ui.adsabs.harvard.edu/abs/2022JOSS....7.4634K} {7, 4634}

\bibitem[\protect\citeauthoryear{{Karamanis}, {Beutler}, {Peacock}, {Nabergoj}
  \& {Seljak}}{{Karamanis} et~al.}{2022b}]{Karamanis2022_MNRAS_516_1644}
{Karamanis} M.,  {Beutler} F.,  {Peacock} J.~A.,  {Nabergoj} D.,   {Seljak} U.,
   2022b, \mn@doi [\mnras] {10.1093/mnras/stac2272}, \href
  {https://ui.adsabs.harvard.edu/abs/2022MNRAS.516.1644K} {516, 1644}

\bibitem[\protect\citeauthoryear{{Kingma} \& {Ba}}{{Kingma} \&
  {Ba}}{2014}]{Kingma2014_arXiv_1412_6980}
{Kingma} D.~P.,  {Ba} J.,  2014, \mn@doi [arXiv e-prints]
  {10.48550/arXiv.1412.6980}, \href
  {https://ui.adsabs.harvard.edu/abs/2014arXiv1412.6980K} {p. arXiv:1412.6980}

\bibitem[\protect\citeauthoryear{{Lange}, {van den Bosch}, {Zentner}, {Wang},
  {Hearin}  \& {Guo}}{{Lange} et~al.}{2019}]{Lange2019_MNRAS_490_1870}
{Lange} J.~U.,  {van den Bosch} F.~C.,  {Zentner} A.~R.,  {Wang} K.,  {Hearin}
  A.~P.,   {Guo} H.,  2019, \mn@doi [\mnras] {10.1093/mnras/stz2664}, \href
  {https://ui.adsabs.harvard.edu/abs/2019MNRAS.490.1870L} {490, 1870}

\bibitem[\protect\citeauthoryear{{Leja} et~al.,}{{Leja}
  et~al.}{2019}]{Leja2019_ApJ_877_140}
{Leja} J.,  et~al., 2019, \mn@doi [\apj] {10.3847/1538-4357/ab1d5a}, \href
  {https://ui.adsabs.harvard.edu/abs/2019ApJ...877..140L} {877, 140}

\bibitem[\protect\citeauthoryear{{Lemos} et~al.,}{{Lemos}
  et~al.}{2023}]{Lemos2023_MNRAS_521_1184}
{Lemos} P.,  et~al., 2023, \mn@doi [\mnras] {10.1093/mnras/stac2786}, \href
  {https://ui.adsabs.harvard.edu/abs/2023MNRAS.521.1184L} {521, 1184}

\bibitem[\protect\citeauthoryear{{Moss}}{{Moss}}{2020}]{Moss2020_MNRAS_496_328}
{Moss} A.,  2020, \mn@doi [\mnras] {10.1093/mnras/staa1469}, \href
  {https://ui.adsabs.harvard.edu/abs/2020MNRAS.496..328M} {496, 328}

\bibitem[\protect\citeauthoryear{Neal}{Neal}{2003}]{Neal2003}
Neal R.~M.,  2003, \mn@doi [The Annals of Statistics] {10.1214/aos/1056562461},
  31, 705

\bibitem[\protect\citeauthoryear{{Nelson} et~al.,}{{Nelson}
  et~al.}{2020}]{Nelson2020_AJ_159_73}
{Nelson} B.~E.,  et~al., 2020, \mn@doi [\aj] {10.3847/1538-3881/ab5190}, \href
  {https://ui.adsabs.harvard.edu/abs/2020AJ....159...73N} {159, 73}

\bibitem[\protect\citeauthoryear{{Pacifici} et~al.,}{{Pacifici}
  et~al.}{2023}]{Pacifici2023_ApJ_944_141}
{Pacifici} C.,  et~al., 2023, \mn@doi [\apj] {10.3847/1538-4357/acacff}, \href
  {https://ui.adsabs.harvard.edu/abs/2023ApJ...944..141P} {944, 141}

\bibitem[\protect\citeauthoryear{{Pedregosa} et~al.,}{{Pedregosa}
  et~al.}{2011}]{Pedregosa2011_JMLR_12_2825}
{Pedregosa} F.,  et~al., 2011, \mn@doi [Journal of Machine Learning Research]
  {10.48550/arXiv.1201.0490}, \href
  {https://ui.adsabs.harvard.edu/abs/2011JMLR...12.2825P} {12, 2825}

\bibitem[\protect\citeauthoryear{{Petigura} et~al.,}{{Petigura}
  et~al.}{2016}]{Petigura2016_ApJ_818_36}
{Petigura} E.~A.,  et~al., 2016, \mn@doi [\apj] {10.3847/0004-637X/818/1/36},
  \href {https://ui.adsabs.harvard.edu/abs/2016ApJ...818...36P} {818, 36}

\bibitem[\protect\citeauthoryear{{Salomone}, {South}, {Drovandi}  \&
  {Kroese}}{{Salomone} et~al.}{2018}]{Salomone2018_arXiv_1805_3924}
{Salomone} R.,  {South} L.~F.,  {Drovandi} C.~C.,   {Kroese} D.~P.,  2018,
  \mn@doi [arXiv e-prints] {10.48550/arXiv.1805.03924}, \href
  {https://ui.adsabs.harvard.edu/abs/2018arXiv180503924S} {p. arXiv:1805.03924}

\bibitem[\protect\citeauthoryear{{Skilling}}{{Skilling}}{2004}]{Skilling2004_AIPC_735_395}
{Skilling} J.,  2004, in {Fischer} R.,  {Preuss} R.,   {Toussaint} U.~V.,  eds,
   American Institute of Physics Conference Series Vol. 735, Bayesian Inference
  and Maximum Entropy Methods in Science and Engineering: 24th International
  Workshop on Bayesian Inference and Maximum Entropy Methods in Science and
  Engineering. pp 395--405, \mn@doi{10.1063/1.1835238}

\bibitem[\protect\citeauthoryear{{Speagle}}{{Speagle}}{2020}]{Speagle2020_MNRAS_493_3132}
{Speagle} J.~S.,  2020, \mn@doi [\mnras] {10.1093/mnras/staa278}, \href
  {https://ui.adsabs.harvard.edu/abs/2020MNRAS.493.3132S} {493, 3132}

\bibitem[\protect\citeauthoryear{{To}, {Rozo}, {Krause}, {Wu}, {Wechsler}  \&
  {Salcedo}}{{To} et~al.}{2023}]{To2023_JCAP_01_016}
{To} C.-H.,  {Rozo} E.,  {Krause} E.,  {Wu} H.-Y.,  {Wechsler} R.~H.,
  {Salcedo} A.~N.,  2023, \mn@doi [\jcap] {10.1088/1475-7516/2023/01/016},
  \href {https://ui.adsabs.harvard.edu/abs/2023JCAP...01..016T} {2023, 016}

\bibitem[\protect\citeauthoryear{{Williams}, {Veitch}  \&
  {Messenger}}{{Williams} et~al.}{2023}]{Williams2023_arXiv_2302_8526}
{Williams} M.~J.,  {Veitch} J.,   {Messenger} C.,  2023, \mn@doi [arXiv
  e-prints] {10.48550/arXiv.2302.08526}, \href
  {https://ui.adsabs.harvard.edu/abs/2023arXiv230208526W} {p. arXiv:2302.08526}

\bibitem[\protect\citeauthoryear{{Zentner}, {Hearin}, {van den Bosch}, {Lange}
  \& {Villarreal}}{{Zentner} et~al.}{2019}]{Zentner2019_MNRAS_485_1196}
{Zentner} A.~R.,  {Hearin} A.,  {van den Bosch} F.~C.,  {Lange} J.~U.,
  {Villarreal} A.~S.,  2019, \mn@doi [\mnras] {10.1093/mnras/stz470}, \href
  {https://ui.adsabs.harvard.edu/abs/2019MNRAS.485.1196Z} {485, 1196}

\bibitem[\protect\citeauthoryear{{Zheng} \& {Guo}}{{Zheng} \&
  {Guo}}{2016}]{Zheng2016_MNRAS_458_4015}
{Zheng} Z.,  {Guo} H.,  2016, \mn@doi [\mnras] {10.1093/mnras/stw523}, \href
  {https://ui.adsabs.harvard.edu/abs/2016MNRAS.458.4015Z} {458, 4015}

\bibitem[\protect\citeauthoryear{{van der Walt}, {Colbert}  \&
  {Varoquaux}}{{van der Walt} et~al.}{2011}]{vanderWalt2011_CSE_13_22}
{van der Walt} S.,  {Colbert} S.~C.,   {Varoquaux} G.,  2011, \mn@doi
  [Computing in Science and Engineering] {10.1109/MCSE.2011.37}, \href
  {https://ui.adsabs.harvard.edu/abs/2011CSE....13b..22V} {13, 22}

\makeatother
\end{thebibliography}

\appendix

\section{Uniform Sampling}
\label{sec:uniform_sampling}

\begin{figure}
    \centering
    \includegraphics{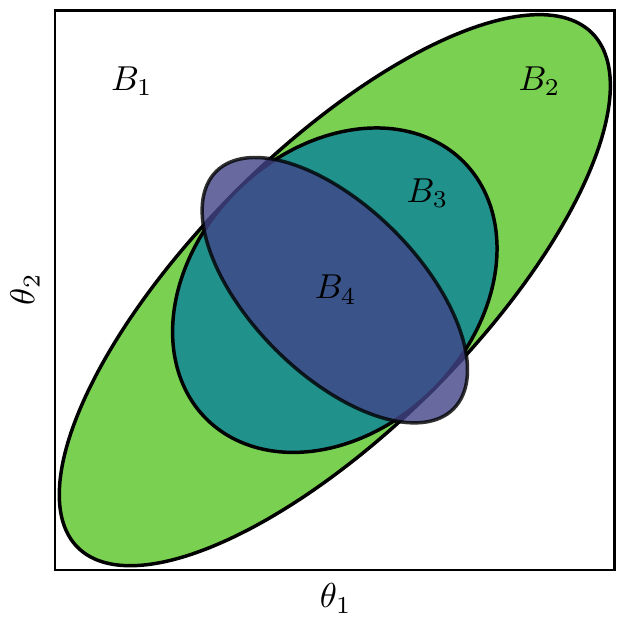}
    \caption{Cartoon depicting potential problems when trying to sample bound uniformly. Even if $B_1$, $B_2$ and $B_3$ were sampled uniformly, points from $B_1$, $B_2$ and $B_3$ falling into $B_4$ do not sample $B_4$ uniformly. See the text for details.}
    \label{fig:non-nested}
\end{figure}

At each iteration $i$ the bound $B_i$ is sampled uniformly with new points. However, points sampled in previous iterations that fall in $B_i$ will not necessarily sample it uniformly. That would be the case if the bounds are strictly nested, i.e., $B_{i + 1}$ is a subset of $B_i$ for all $i$. However, this is not generally the case, as illustrated in the cartoon in Fig.~\ref{fig:non-nested}. In this example, the INS algorithm has constructed four bounds and sampled the first three of them. However, $B_4$ is not a subset of $B_3$ and instead also overlaps, for example, with $B_1 \setminus B_2$. As a result, the volumes $B_4 \cap B_3$ and $B_4 \cap (B_1 \setminus B_2)$ will be sampled at different densities. As described in the following, this kind of issue can be resolved straightforwardly. Let us assume we populate bound $B_l$. We can define the $l-1$ volumes $S_{i|l-1}$ as the parts of the prior space that were shells if only considering the first $l-1$ bounds, i.e.
\begin{equation}
    S_{i|l-1} = B_i \setminus \left( \cup_{k=i+1}^{l-1} B_k \right) \, .
\end{equation}
We initially remove all points that fall into $B_l$ and put them into a ``replacement set''. For points in the replacement set, we can determine which previous shell $S_{i|l-1}$ it belongs to. We then start to sample from $B_l$ uniformly. If a newly sampled point belongs to the same shell $S_{i|l-1}$ as a point in the replacement set, we disregard it and replace it with the one from the replacement set. The likelihood of this point from the replacement set is already known and does not need to be re-evaluated. If, on the other hand, no point in the replacement set belongs to the same shell $S_{i|l-1}$ as the newly drawn point, we evaluate the newly drawn point.

\section{Arbitrary priors}
\label{sec:arbitrary_prior}

In certain situations, the prior $\pi(\mathbf{\Theta})$ may not be easily separable and the approach introduced by \cite{Alsing2021_MNRAS_505_95} not be desirable or feasible. In this case, one can work with an alternative flat, normalized and separable prior $\tilde{\pi}$ over some volume that fully covers the original prior. One can now define an alternative likelihood that incorporates the original prior, i.e. $\tilde{\mathcal{L}}(\mathbf{\Theta}) = \mathcal{L}(\mathbf{\Theta}) \pi(\mathbf{\Theta})$. The evidence of this alternative prior and likelihood is $\tilde{\mathcal{Z}} = \mathcal{Z} V^{-1}$, where $V$ is the volume covered by $\tilde{\pi}$. Similarly, samples drawn from this alternative prior and likelihood combination are statistically the same as those drawn from the original combination. However, as discussed in \citep{Feroz2009_MNRAS_398_1601} may increase the overall runtime since a larger part of parameter space has to be explored.

\label{lastpage}

\end{document}